\documentclass [11pt,a4paper] {article}

\usepackage[T1]{fontenc}
\usepackage{lmodern}

\usepackage[cp1252]{inputenc}
\usepackage{amssymb}
\usepackage{amsmath}
\usepackage{amsfonts,amssymb}
\usepackage[dvips]{graphicx}
\usepackage{bbm}
\usepackage{enumerate}

\usepackage[shortlabels]{enumitem}

\usepackage{amsthm}
\usepackage{cancel}

\usepackage{centernot}
\usepackage{mathrsfs}

\usepackage{wasysym}

\DeclareMathAlphabet{\mathpzc}{OT1}{pzc}{m}{it}

\def\SmallColSep{\setlength{\arraycolsep}{1pt}}

\newcommand*\rfrac[2]{{}^{#1}\!\!/\!_{#2}}

\newcommand{\notle}{\mathrel{{\ooalign{\hidewidth$\not\phantom{"}$\hidewidth\cr$\le$}}}}

\begin{document}

\title{Physics in a finite geometry}

\author{Arkady Bolotin\footnote{$Email: arkadyv@bgu.ac.il$\vspace{5pt}} \\ \emph{Ben-Gurion University of the Negev, Beersheba (Israel)}}

\maketitle

\begin{abstract}\noindent The stipulation that no measurable quantity could have an infinite value is indispensable in physics. At the same time, in mathematics, the possibility of considering an infinite procedure as a whole is usually taken for granted. However, not only does such possibility run counter to computational feasibleness, but it also leads to the most serious problem in modern physics, to wit, the emergence of infinities in calculated physical quantities. Particularly, having agreed on the axiom of infinity for set theory – the backbone of the theoretical foundations of calculus integrated in every branch of physics – one could no longer rule out the existence of a classical field theory which is not quantizable, let alone renormalizable. By contrast, the present paper shows that negating the axiom of infinity results in physics acting in a finite geometry where it is ensured that all classical field theories are quantizable.\bigskip\bigskip

\bigskip\bigskip
\end{abstract}

\section{Introduction}  

\noindent According to reductionism, one theory, say $\texttt{T}$, can be reducible to another, e.g., $\texttt{W}$, which is more fundamental than $\texttt{T}$. This means that either all truths of $\texttt{T}$ can be translated into the language of $\texttt{W}$, or the laws of $\texttt{T}$ can be derived from those of $\texttt{W}$, or all the observations explained by $\texttt{T}$ can be explained by $\texttt{W}$ \cite{Ney}.\bigskip

\noindent The goal of reductionism is to provide some sense in which science may become more unified \cite{Ruse}. Herewith, every unification step, for example, the step from $\texttt{T}$ to $\texttt{W}$, leads one level up closer to the most fundamental theory (so called \emph{theory of everything}, or TOE for short) that unifies all theories in physics and other sciences (on condition that those sciences are all reducible to physics) and also fully explains and links together all physical aspects of the universe \cite{Weinberg}.\bigskip

\noindent Therefore, finding a TOE has become one of major tasks in physics. Remarkably, despite the intense search for the ultimate physical theory, neither a unification of strong and electroweak forces, nor a unification of quantum theory and general relativity had been achieved \cite{Wang, Hodge}.\bigskip

\noindent In the pursuit of TOE, one of the toughest questions yet to be answered is how to quantize an arbitrary classical field theory, that is, to merge it with quantum mechanics.\bigskip

\noindent In line with the reductionist approach, the transition from a classical field to a quantum operator field should be analogous to the promotion of a classical harmonic oscillator to a quantum harmonic oscillator. At the same time, such a quantization entails the decomposition of solutions to the classical equations of motion into formal power series known as perturbation series. On the assumption that the axiom of infinity of Zermelo-Fraenkel set theory holds true in any physical formalism, the perturbation series give rise to divergent sums implying that some terms of these series must be divergent. Because of that, “counterterms” cancelling the divergent terms are required in order to obtain a finite result \cite{Shifman}.\bigskip

\noindent Therewithal, the major concern is the number of the counterterms: A quantum counterpart of a classical filed theory will make any sense (i.e., will be well-defined) if that number is finite. Only then the classical filed theory is said to be \emph{quantizable} and \emph{renormalizable}. In case the number of the counterterms ends up being infinite, the classical theory will be quantizable but nonrenormalizable \cite{Kaku}. Thus, general relativity is a quantizable nonrenormalizable classical theory of gravitational field \cite{Rovelli}.\bigskip

\noindent But in what follows, it will be shown that in the context of a theory of sets with the negation of the axiom of infinity, any classical filed theory can be quantized in a way where its quantum counterpart does not contain divergences. This suggests that the answer to the question whether given physical theories can be unified is determined by the set-theoretic axioms underlaying formalisms of those theories.\bigskip

\section{Tarski's material adequacy condition}  

\noindent Let us start by pointing out that any theory $\texttt{T}$ can be formulated as a collection of sentences, i.e., words or groups of words, in some ordinary language and some formal language.\bigskip

\noindent Consider a sentence of a theory $\texttt{T}$ in an ordinary language. Let us refer to it as a sentence $S$. Suppose that $S$ is such that it makes sense to ask if $S$ is true in $\texttt{T}$. Let $\Phi_S$ denote a formal representation of $S$, i.e., a description (a copy) of $S$ in words (well-formed formulas, WFFs for short) of a formal language of the theory $\texttt{T}$ referred to as $\mathcal{L}(\texttt{T})$. Since $S$ is capable of being true or false, $\Phi_S$ can be given a meaning (in more general terms, it can be brought into a representational relation to the world).\bigskip 

\noindent By way of illustration, let the sentence $S$ of \emph{number theory} having the truth value of ``true'' be as follows: ``The set of non-negative even integers is not empty''. Its formal representation is $\Phi_S \mathrel{\mathop:}= \{x \in \mathbb{Z}|(2|x) \sqcap x \ge 0\}\neq \{x|x \neq x\}$, where $2|x$ indicates that 2 divides $x$ and the symbol $\sqcap$ stands for the logical connective ``and''. Needless to say, $\Phi_S$ is meaningful.\bigskip

\noindent The sentence $S$ can be categorized as being belonged to \emph{the object language} of the theory $\texttt{T}$. The said language may differ from an ordinary language by having concepts lacking in the latter (for example, the concept of \emph{empty set} cannot be constructed from ordinary linguistic notions such as ``nothingness'' or ``nonexistence''). Correspondingly, the formal language $\mathcal{L}(\texttt{T})$ can be regarded as \emph{the metalanguage} with respect to the object language of the theory $\texttt{T}$ \cite{Kirkham}.\bigskip

\noindent Then, in line with Tarski's material adequacy condition \cite{Horsten, Hodges} one can declare that $S$ will be true in $\texttt{T}$ if and only if $\Phi_S$ is true in $\mathcal{L}(\texttt{T})$. Symbolically, this can be presented as the T-sentence:\smallskip

\begin{equation}  
   S
   \text{ is true in }
   \texttt{T}
   \iff
   \Phi_S
   \text{ is true in }
   \mathcal{L}(\texttt{T})
   \;\;\;\;  ,
\end{equation}
\smallskip

\noindent where $\iff$ stands for ``if and only if''.\bigskip

\noindent Assume that for all sentences $S$ in an object language of $\texttt{T}$, those $S$ are in one-to-one correspondence with their formal representations $\Phi_S$. In that case, $\texttt{T}$ is \emph{a formal theory}, i.e., a theory whose every sensible sentence in the object language can be described by WFFs of the formal language using some former grammar. Obvious examples of formal theories are logic and mathematics.\bigskip

\noindent The function from $\Sigma_S$, the set of sentences $S$ in the object language of $\texttt{T}$, to $\Sigma_{\Phi_S}$, the set of WFFs in the formal language $\mathcal{L}(\texttt{T})$, denoted by\smallskip

\begin{equation}  
   \begingroup\SmallColSep
   \begin{array}{l l l l}
   f\!\!: & \,\Sigma_S & \to        & \Sigma_{\Phi_S} \\
          & \,S             & \mapsto & \Phi_S
   \end{array}
   \endgroup
   \;\;\;\;  ,
\end{equation}
\smallskip

\noindent can be called \emph{a formalization of a theory} $\texttt{T}$ because it determines how far the object language of $\texttt{T}$ can be formalized by means of its formal language $\mathcal{L}(\texttt{T})$. E.g., in the case of a formal theory, the above function is a bijection; so, such a theory is fully formalized (for instance, logic and mathematics are fully formalized theories).\bigskip

\noindent In contrast, let $f\!\!: \Sigma_S \not \to \Sigma_{\Phi_S}$ indicate that $f$ is a multivalued or partial function. In that case, there is at least one sensible sentence in the object language of a theory $\texttt{T}$ that has either no formal representation or more than one. Accordingly, that theory can be classified as \emph{partially formalized}.\bigskip

\noindent An instance of a partially formalized theory is quantum mechanics ($\texttt{QM}$), for which the T-sentence takes the form\smallskip

\begin{equation}  
   S
   \text{ is true in }
   \texttt{QM}
   \iff
   \Phi_S
   \text{ is true in }
   \mathcal{L}(\texttt{QM})
   \;\;\;\;  ,
\end{equation}
\smallskip

\noindent where $S$ denotes a sensible sentence in the object language of $\texttt{QM}$, as for example ``The spin of a spin-$\rfrac{1}{2}$ particle is \emph{up} along the $\overrightarrow{\!\! z}$ axis'', and $\Phi_S$ is the copy of $S$ in the formal language of quantum mechanics $\mathcal{L}(\texttt{QM})$ that has the axioms of Hilbert space theory at its core \cite{Edwards}. To be precise, the axiomatization of $\texttt{QM}$ can be presented by $\texttt{Hil}$, the collection of the axioms of Hilbert space theory, in conjunction with $\textbf{\texttt{X}}_{\texttt{QM}}$, the axioms of $\texttt{QM}$ that are absent in $\texttt{Hil}$ (such as Born Rule). The last implies that $\mathcal{L}(\texttt{QM})$ can be displayed as $\mathcal{L}(\texttt{Hil} \dotplus \textbf{\texttt{X}}_{\texttt{QM}})$ where the symbol $\dotplus$ stands for “in conjunction with”. Therefore, $\Phi_S \in \mathcal{L}(\texttt{Hil} \dotplus \textbf{\texttt{X}}_{\texttt{QM}})$.\bigskip

\noindent To illustrate why $\texttt{QM}$ can only be partially formalized, let us take two sensible sentences $S_1$ and $S_2$ of the object language of $\texttt{QM}$ involving incompatible quantum properties (such as the spin of a spin-$\rfrac{1}{2}$ particle along the different axes) and join them together to form the compound sentence $S_1 \sqcap S_2$. At that juncture, do realize that the said sentence does not have a unique representation in $\mathcal{L}(\texttt{Hil} \dotplus \textbf{\texttt{X}}_{\texttt{QM}})$. That is, when $S_1$ and $S_2$ are incompatible, there is no unique relation from $S_1 \sqcap S_2$ to a closed linear subspace of a Hilbert space endowed with a countably additive probability measure. The same holds true for the compound sentence $S_1 \sqcup S_2$ where $\sqcup$ denotes ``or'' (more detail can be found in \cite{Bolotin}).\bigskip

\noindent The function that undoes the mapping $f\!\!: \Sigma_S \to \Sigma_{\Phi_S}$ is its inverse, $f^{-1}\!\!: \Sigma_{\Phi_S} \to \Sigma_S$, which sends every WFF of the formal language to a sentence of the object language. Due to this, one can say that the function $f^{-1}\!\!: \Sigma_{\Phi_S} \to \Sigma_S$ determines \emph{a semantic description} of the formal language; in other words, it gives an interpretation of the formalism of a theory.\bigskip

\noindent Clearly, $f^{-1}$ exists if and only if $f$ is bijective. This means that the unique interpretation of the formalism is possible only if a theory is fully formalized. Consequently, for a theory that is only partially formalized (such as $\texttt{QM}$), more than one interpretation of its formalism may be possible.\bigskip

\section{The argument against reductionism}  

\noindent Intuitively, the more powerful a theory is, the greater the variety and quantity of ideas can be expressed through its formal language. Put differently, the power of a theory is determined by the expressiveness (\emph{expressive power}) of its formal language. For this reason, when comparing two theories $\texttt{T}$  and $\texttt{W}$ with an eye to decide which of them is reducible to the other, it makes sense to compare the expressive powers of their formal languages $\mathcal{L}(\texttt{T})$ and $\mathcal{L}(\texttt{W})$.\bigskip

\noindent For ease of reference, let us denote WFFs of $\mathcal{L}(\texttt{T})$ and $\mathcal{L}(\texttt{W})$ by $\Phi$ and $\Psi$ respectively. Consider $\Sigma_{\Phi}$ and $\Sigma_{\Psi}$, the sets of $\Phi$ and $\Psi$ in the order given. Let us assume that there is a way of matching $\Phi$ with $\Psi$. In that case, two options are possible:\smallskip

\begin{enumerate}[leftmargin=1.5cm]
\item The sets $\Sigma_{\Phi}$ and $\Sigma_{\Psi}$ have the same cardinality:\vspace{-6pt}

\begin{equation} 
   \mathrm{card}\left(\Sigma_{\Phi}\right) = \mathrm{card}\left(\Sigma_{\Psi}\right)
   \;\;\;\;  .
\end{equation}

\item The sets $\Sigma_{\Phi}$ and $\Sigma_{\Psi}$ have unequal cardinalities, specifically, one of them has the cardinality strictly less than the cardinality of the other:\vspace{-6pt}

\begin{equation} \label{EXPR} 
   \mathrm{card}\left(\Sigma_{\Phi}\right) \lessgtr \mathrm{card}\left(\Sigma_{\Psi}\right)
   \;\;\;\;  .
\end{equation}

\end{enumerate}
\bigskip

\noindent The first option indicates that there is a bijection from $\Sigma_{\Phi}$ to $\Sigma_{\Psi}$, i.e., a function from WFFs in one formal language to WFFs in another that is both injective and surjective:\smallskip

\begin{equation}  
   \begingroup\SmallColSep
   \begin{array}{l c l c}
   f\!\!: & \,\Sigma_{\Phi} & \to          & \Sigma_{\Psi} \\
          & \,\Phi                & \mapsto   & \Psi
   \end{array}
   \endgroup
   \;\;\;\;  .
\end{equation}
\smallskip

\noindent This means that everything that can be expressed through the language $\mathcal{L}(\texttt{T})$ can also be expressed through the language $\mathcal{L}(\texttt{W})$, and vice versa. Ergo, $\mathcal{L}(\texttt{T})$ and $\mathcal{L}(\texttt{W})$ have the same expressive power. Such languages can be said to be \emph{synonymous} or \emph{isomorphic}. To give an instance, the formal languages over square integrable functions $\Psi \in L^{2}(\mathbb{R},dx)$ (i.e., wave functions), square summable sequences $l_{2}=\{c_{n} \in \mathbb{C}|\sum_{n=0}^{\infty}|c_{n}|^{2}\}$ (i.e., sums of the absolute squares of the matrix elements), and vector spaces $\mathcal{H}$ with scalar product $\langle , \rangle \ge 0$, denoted by $\mathcal{L}(L^2)$, $\mathcal{L}(l_2)$, and $\mathcal{L}(\mathcal{H})$ respectively, are isomorphic.\bigskip

\noindent The second option reveals that there is an injective function, but no bijective one, i.e., either $i\!\!: \Sigma_{\Phi} \to \Sigma_{\Psi}$ or $i\!\!: \Sigma_{\Psi} \to \Sigma_{\Phi}$. As a result, every WFF of the domain can be presented as a WFF of the codomain (with or without transformations attributable to the difference in grammar). But together with that, there are some WFFs in the codomain that do not have counterparts in the domain. It means that the language, in which WFFs of the codomain are written, is more expressive than that of the domain. The consequence of this is that one of $\mathcal{L}(\texttt{T})$ and $\mathcal{L}(\texttt{W})$ is a metalanguage and the other is an object language. Or, in terms of reductionism, one member of the set with $\texttt{T}$ and $\texttt{W}$ is \emph{the base theory} and the other is \emph{the target theory} \cite{Ney}.\bigskip

\noindent Keep in mind that every axiom is a WFF (but not every WFF is an axiom). According to (\ref{EXPR}), this implies that adding a new axiom to a formal language increases its expressive power.\bigskip

\noindent In case that there is no way of matching axioms of $\mathcal{L}(\texttt{T})$ and $\mathcal{L}(\texttt{W})$, those languages are clear distinct, but neither is more expressive than the other. Consequently, the pair cannot be an object language and a metalanguage. Said otherwise, $\mathcal{L}(\texttt{T})$ and $\mathcal{L}(\texttt{W})$ cannot be in a hierarchical relation, and so the theories $\texttt{T}$ and $\texttt{W}$ cannot include one another.\bigskip

\noindent From the point of view of reductionism, this means that whenever the axioms of $\mathcal{L}(\texttt{T})$ and $\mathcal{L}(\texttt{W})$ cannot be matched, the theories $\texttt{T}$ and $\texttt{W}$ cannot be reducible among themselves. Clearly, this represents the limitation of reductionism.\bigskip

\section{Hilbert space theory as an axiomatic set theory}  

\noindent Let us assess the expressive power of $\mathcal{L}(\texttt{Hil})$. A way to do this is to present Hilbert space theory in the form of some set theory whose axiomatization is given by the collection of axioms $\texttt{ST}_{\texttt{Hil}}$. On condition that $\texttt{Hil}$ is isomorphic to $\texttt{ST}_{\texttt{Hil}}$, one can set the latter against $\texttt{ZF}$, the collection of axioms of Zermelo-Fraenkel set theory. Following this, one is allowed to compare the power of Hilbert space theory with that of $\texttt{ZF}$, the most common foundation of mathematics.\bigskip

\noindent In accordance with an axiomatic set theory, sets are \emph{undefined terms} \cite{Holmes}. This means that in the definition ``Sets are collections of things'', it is not necessary to know what ``things'' are which constitute ``sets''. That is to say, an axiomatic theory of sets makes no distinction between \emph{a set} and \emph{a mathematical structure}, which is a set possessing some additional features \cite{MacLane}. An example of such a structure is \emph{a vector space} (a.k.a. linear space), i.e., a set of objects (called vectors) being endowed with certain operations (known as vector addition and scalar multiplication) that satisfy some requirements (called vector axioms) \cite{Mirsky}.\bigskip

\noindent On the grounds of this, one may establish the axiomatic system $\texttt{ST}_{\texttt{Hil}}$ by replacing ``sets'' with ``vector spaces'' in the $\texttt{ZF}$ axioms. Let us demonstrate how this can be accomplished.\bigskip

\subsection{The axiom of extension \textbf{\texttt{Ext}}} 

\noindent In the object language of $\texttt{ZF}$, the axiom \textbf{\texttt{Ext}} reads (here and henceforth, the $\texttt{ZF}$ axioms are given in accordance with \cite{Karel}): ``Two sets $X$ and $Y$ are equal if and only if they have the same elements $z$''. Its representation in the formal language of $\texttt{ZF}$ is:\smallskip

\begin{equation}  
   \textbf{\texttt{Ext}}
   \mathrel{\mathop:}=
   \forall X \forall Y
   \left[
      \forall z \left( z \in X \iff z \in Y\right)
      \to X = Y
   \right]
   \;\;\;\;  .
\end{equation}
\smallskip

\noindent Putting ``vector spaces'' in place of ``sets'', one gets this axiom expressed in the object language of Hilbert space theory as follows: ``Two vector spaces $\mathcal{H}_1$ and $\mathcal{H}_2$ are equal if and only if they have the same vectors $|\Psi\rangle$”. Using WFFs of $\mathcal{L}(\texttt{Hil})$, it takes the form:\smallskip

\begin{equation}  
   \textbf{\texttt{Ext}}
   \mathrel{\mathop:}=
   \forall \mathcal{H}_1 \forall \mathcal{H}_2
   \left[
      \forall |\Psi\rangle \left( |\Psi\rangle \in \mathcal{H}_1 \iff |\Psi\rangle \in \mathcal{H}_2\right)
      \to \mathcal{H}_1 = \mathcal{H}_2
   \right]
   \;\;\;\;  .
\end{equation}

\subsection{The axiom schema of specification (separation) \textbf{\texttt{Spec}}} 

\noindent In the object language of $\texttt{ZF}$, this axiom schema states that a subset $Y$ of a set $X$ \emph{always exits}. Or, at greater length, given a set $X$ and a predicate $\Phi$, there always exits a subset $Y$ of $X$ whose members $x$ are precisely the members of $X$ that satisfy $\Phi$. In $\mathcal{L}(\texttt{ZF})$, this may be written as:\smallskip

\begin{equation}  
   \textbf{\texttt{Spec}}
   \mathrel{\mathop:}=
   \forall w_1,\dots\,w_n
   \forall X \exists Y \forall x
   \left[
      x \in Y \iff \left( x \in X \sqcap \Phi\left(x,w_1,\dots,w_n,X\right) \right)
    \right]
   \;\;\;\;  ,
\end{equation}
\smallskip

\noindent where $\Phi$ is any formula in $\mathcal{L}(\texttt{ZF})$ with free variables containing $x,w_1,\dots,w_n,X$.\bigskip

\noindent Let us assume that the predicate $\Phi$ is false. Since it is postulated that $Y$ always exists, $Y$ must be such that no element is a member of it. In this way, \textbf{\texttt{Spec}} proves the existence of an empty set.\bigskip

\noindent By replacing ``set'' with ``vector space'', the axiom schema of specification translates as: Given a vector space $\mathcal{H}$ and a predicate $\Phi$, there \emph{may possibly} exist a subspace $\mathcal{V}$ of $\mathcal{H}$, whose vectors $|\Psi\rangle$ are precisely the members of $\mathcal{H}$ that satisfy $\Phi$. Such possibility becomes real if and only if $\mathcal{V}$ is a vector space under the same operations that makes a set of vectors $\mathcal{H}$ into a vector space. In $\mathcal{L}(\texttt{Hil})$, this translation can be presented by the expression:\smallskip

\begin{equation}  
   \textbf{\texttt{Spec}}_{\diamond}
   \mathrel{\mathop:}=
   \forall w_1,\dots\,w_n
   \forall \mathcal{H} \Diamond\exists \mathcal{V} \forall |\Psi\rangle
   \Big[
      |\Psi\rangle \in \mathcal{V} \iff \left( |\Psi\rangle \in \mathcal{H} \sqcap \Phi\left(|\Psi\rangle,w_1,\dots,w_n,\mathcal{H}\right) \right)
    \Big]
   \;\;\;\;  ,
\end{equation}
\smallskip

\noindent where the symbol $\Diamond$ denotes the modal operator \emph{Possibly} (so that $\Diamond\exists \mathcal{V}$ reads ``It is possible that $\mathcal{V}$ exists'') and $\Phi$ is any formula in $\mathcal{L}(\texttt{H})$ with free variables containing $|\Psi\rangle,w_1,\dots,w_n,\mathcal{H}$.\bigskip

\noindent Unlike \textbf{\texttt{Spec}}, the axiom $\textbf{\texttt{Spec}}_{\diamond}$ does not permit the existence of a collection of things such that no element is a member of it. To be a vector space, a set of vectors $|\Psi\rangle$ must include at least the zero vector, 0; as a result, if the predicate $\Phi$ is false, $\mathcal{V}$ does not exist. Thus, the notion of \emph{empty set} is not translatable into the notion of \emph{empty vector space}.\bigskip

\noindent Take notice that the axiom \textbf{\texttt{Spec}} in conjunction with the negation of the axiom of empty set \textbf{\texttt{Empty}} (asserting the existence of a set with no elements) leads to the axiom $\textbf{\texttt{Spec}}_{\diamond}$ as a logical consequence. Symbolically, this can be presented as\smallskip

\begin{equation}  
   \textbf{\texttt{Spec}}
   \dotplus
   \neg\textbf{\texttt{Empty}}
   \vdash
   \textbf{\texttt{Spec}}_{\diamond}
   \;\;\;\;  .
\end{equation}

\subsection{The axiom of regularity (a.k.a. the axiom of foundation) \textbf{\texttt{Reg}}} 

\noindent The following statement is the axiom of regularity of $\texttt{ZF}$ set theory: ``Every non-empty set $X$ contains an element that is disjoint from $X$''. In the formal language $\mathcal{L}(\texttt{ZF})$, this statement is:\smallskip

\begin{equation}  
   \textbf{\texttt{Reg}}
   \mathrel{\mathop:}=
   \forall X
   \left[
      X \neq \varnothing \to \exists Y\left(Y \in X \right) \sqcap \left(Y \cap X = \varnothing \right)
    \right]
   \;\;\;\;  ,
\end{equation}
\smallskip

\noindent where $\varnothing$ stands for the empty set.\bigskip

\noindent Let $X$ and $Y$ be the singleton $\{A\}$ and the set $A$ respectively. From \textbf{\texttt{Reg}} it follows then that $A\in\{A\}$ and $A\cap\{A\}=\varnothing$. Therefore, $A\notin A$, i.e., no set is element of itself. The last implies that a singleton is necessarily distinct from the element it contains.\bigskip

\noindent Do not forget that in compliance with $\textbf{\texttt{Spec}}_{\diamond}$, there is no empty vector space. Consequently, any vector space $\mathcal{H}$ must have at least one element, specifically, the trivial (or zero) vector space $\{0\}$ whose intersection with $\mathcal{H}$ is the trivial space again. Keeping this in mind, consider the propositional formula $P(\mathcal{V})$ with one free variable $\mathcal{V}$, namely,\smallskip

\begin{equation}  
   P(\mathcal{V})
   =
   \left(\mathcal{V} \in \mathcal{H} \right) \sqcap \left(\mathcal{V} \cap \mathcal{H} = \{0\} \right)
   \;\;\;\;  ,
\end{equation}
\smallskip

\noindent where $\mathcal{H}$, an arbitrary vector space, is a bound variable. Since this formula is known to be true for $\mathcal{V} = \{0\}$, one can evoke existential introduction along with universal generalization, viz.,

\begin{equation}  
   P(\{0\})
   \to
   \forall \mathcal{H} \exists \mathcal{V} P(\mathcal{V})
   \;\;\;\;  ,
\end{equation}
\smallskip

\noindent to find that for all vector spaces $\mathcal{H}$, there is an element that is \emph{almost} disjoint from $\mathcal{H}$ (meaning that given any $\mathcal{H}$, its intersection with such an element only contains the zero vector space $\{0\}$). Hence, the axiom \textbf{\texttt{Reg}} translates into the formal language $\mathcal{L}(\texttt{Hil})$ as follows:\smallskip

\begin{equation}  
   \textbf{\texttt{Reg}}_{\,\texttt{almost}}
   \mathrel{\mathop:}=
   \forall \mathcal{H}
   \Big[
      \mathcal{H} \neq \{0\} \to \exists \mathcal{V}\left(\mathcal{V} \in \mathcal{H} \right) \sqcap \left(\mathcal{V} \cap \mathcal{H} = \{0\} \right)
    \Big]
   \;\;\;\;  .
\end{equation}
\smallskip

\noindent Again, observe that\smallskip

\begin{equation}  
   \textbf{\texttt{Reg}}
   \dotplus
   \neg\textbf{\texttt{Empty}}
   \vdash
   \textbf{\texttt{Reg}}_{\,\texttt{almost}}
   \;\;\;\;  .
\end{equation}

\subsection{The axiom of power set \textbf{\texttt{PS}}} 

\noindent In the object language of $\texttt{ZF}$, the axiom \textbf{\texttt{PS}} says: ``For every set $X$, there is a set $P(X)$ consisting precisely of all the subsets of $X$''. Its formal representation in $\mathcal{L}(\texttt{ZF})$ is\smallskip

\begin{equation}  
   \textbf{\texttt{PS}}
   \mathrel{\mathop:}=
   \forall X \exists P(X) \forall Z
   \left[
      Z \in P(X) \iff \forall x \left(x \in Z \to x \in X \right)
    \right]
   \;\;\;\;  .
\end{equation}
\smallskip

\noindent Let us point out that the set $P(X)$ (called \emph{the power set of $X$}) can be ordered via subset inclusion to obtain a lattice $L(X)$ bounded by $X$ itself and the empty set $\varnothing$.\bigskip

\noindent In the theory of vector spaces, the counterpart of $L(X)$ is \emph{the vector lattice} $L(\mathcal{H})$, i.e., the set of all closed linear subspaces of a Hilbert space $\mathcal{H}$ ordered through subset inclusion and bounded by $\mathcal{H}$ along with the zero vector space $\{0\}$. Withal, the vector lattice $L(\mathcal{H})$ is a vector space.\bigskip

\noindent Hence, when translated from the language of $\texttt{ZF}$ into the language of Hilbert space theory, the axiom \textbf{\texttt{PS}} becomes as follows: ``For every vector space $\mathcal{H}$, there is a vector space $\mathcal{L}(\mathcal{H})$ consisting precisely of all subspaces of $\mathcal{H}$''. Symbolically, i.e., in WFFs of $\mathcal{L}(\texttt{Hil})$:\smallskip

\begin{equation}  
   \textbf{\texttt{PS}}
   \mathrel{\mathop:}=
   \forall \mathcal{H} \exists L(\mathcal{H}) \forall \mathcal{V}
   \Big[
      \mathcal{V} \in L(\mathcal{H}) \iff \forall |\Psi\rangle \left(|\Psi\rangle \in \mathcal{V} \to |\Psi\rangle \in \mathcal{H} \right)
    \Big]
   \;\;\;\;  .
\end{equation}

\subsection{The axiom of union \textbf{\texttt{Uni}}} 

\noindent In words of the objective language of $\texttt{ZF}$, the axiom $\textbf{\texttt{Uni}}$ states that for any set of sets $S(X)$ there is a set $X$ containing every element that is a member of some member of $S(X)$. In WFFs of $\mathcal{L}(\texttt{ZF})$ this means:\smallskip

\begin{equation}  
   \textbf{\texttt{Uni}}
   \mathrel{\mathop:}=
   \forall S(X) \exists X \forall Y \forall x
   \left[
      \left(x \in Y \sqcap Y \in S(X) \right) \to x \in X
    \right]
   \;\;\;\;  .
\end{equation}
\smallskip

\noindent Recall that ``set of sets'' translates into the language of Hilbert space theory as ``vector lattice''. In addition, notice that if a set of vectors $B$ is a basis for the vector lattice $L(B)$, then it is also a basis for the vector space $\mathrm{span}(B)$, i.e., the linear span of vectors belonging to $B$. In view of that, $\mathrm{span}(B)$ is the intersection of all subspaces of $L(B)$.\bigskip

\noindent Thus, in terms of the objective language of Hilbert space theory, the axiom $\textbf{\texttt{Uni}}$ says that for any vector lattice $L(B)$, there is a vector space $\mathrm{span}(B)$ containing every element that is a member of some member of $L(B)$. In WFFs of $\mathcal{L}(\texttt{Hil})$:\smallskip

\begin{equation}  
   \textbf{\texttt{Uni}}
   \mathrel{\mathop:}=
   \forall L(B) \exists \,\mathrm{span}(B) \forall \,\mathcal{V} \,\forall |\Psi\rangle
   \Big[
      \left(|\Psi\rangle \in \mathcal{V} \sqcap \mathcal{V} \in L(B) \right) \to |\Psi\rangle \in \mathrm{span}(B)
    \Big]
   \;\;\;\;  .
\end{equation}
\smallskip

\noindent Essentially, the axiom $\textbf{\texttt{Uni}}$ allows one to unpack a vector lattice $L(B)$ and thus create a “flatter” vector space, $\mathrm{span}(B)$.\bigskip

\subsection{The axiom schema of replacement \textbf{\texttt{Rep}}} 

\noindent The axiom schema $\textbf{\texttt{Rep}}$ states that if $F$ is \emph{a definable function} (i.e., one that can be defined by some formula with or without parameters), the image of a set under $F$ will also fall inside a set.\bigskip

\noindent In the object language of Hilbert space theory, the replacement schema $\textbf{\texttt{Rep}}$ can be translated as follows: Let a binary relation $T$ be defined as functional, serial, and linear. Then the image of a vector space under $T$ will also be a vector space.\bigskip

\subsection{The axiom of infinity \textbf{\texttt{Inf}}} 

\noindent Consistent with the axiom of regularity $\textbf{\texttt{Reg}}$, a set $X$ is an element of the singleton set $\{X\}$; for this reason, the union of $X$ with $\{X\}$ must include $X$. Calling this union \emph{the successor of $X$} and denoting it by $X^{+}$, namely,\smallskip

\begin{equation}  
   X^{+}
   \mathrel{\mathop:}=
   \left( X \cup \{X\} \right) \ni X
   \;\;\;\;  ,
\end{equation}
\smallskip

\noindent the axiom $\textbf{\texttt{Inf}}$ can be set forth as\smallskip

\begin{equation}  
   \textbf{\texttt{Inf}}
   \mathrel{\mathop:}=
   \exists \,\mathbf{I}
      \left[
         \varnothing \in \mathbf{I} \sqcap \forall X \in \mathbf{I} \left(X^{+} \in \mathbf{I} \right)
      \right]
   \;\;\;\;  .
\end{equation}
\smallskip

\noindent In words of the objective language of $\texttt{ZF}$, the above restates as follows: There is a set $\mathbf{I}$ which has the empty set $\varnothing$ as an element, and whenever some set $X$ is a member of $\mathbf{I}$, the successor of $X$ is also a member of $\mathbf{I}$.\bigskip

\noindent Suppose that some specific set $X$ is in $\mathbf{I}$. Then, as stated by $\textbf{\texttt{Inf}}$, its successor, $X^{+}$, is also in $\mathbf{I}$. But since $X^{+}$ is in $\mathbf{I}$, its successor, ${X^{+}}^{+}$, must be in $\mathbf{I}$ as well. Continuing similarly, one gets the endless sequence of the successors of $X$\smallskip

\begin{equation}  
   X
   \in
   X^{+}
   \in
   {X^{+}}^{+}
   \in
   {{X^{+}}^{+}}^{+}
   \in
   \dots
   \;\;\;\;  ,
\end{equation}
\smallskip

\noindent each belonging to $\mathbf{I}$. As every set in this sequence differs from the one that follows next (otherwise, $X^{+}={X}$ and so $X \ni X$, i.e., any set in the sequence would have been an element of itself), $\mathbf{I}$ must include infinitely many sets.\bigskip

\noindent When translated from $\mathcal{L}(\texttt{ZF})$ into $\mathcal{L}(\texttt{Hil})$, the axiom $\textbf{\texttt{Inf}}$ becomes\smallskip

\begin{equation} 
   \textbf{\texttt{Inf}}_{\mathcal{L}(\texttt{Hil})}
   \mathrel{\mathop:}=
   \exists \,\mathscr{I}
      \left[
         \forall \mathcal{H} \in \mathscr{I} \left(\mathcal{H}^{+} \in \mathscr{I} \right)
      \right]
   \;\;\;\;   
\end{equation}
\smallskip

\noindent that reads: There is a vector space $\mathscr{I}$ such that whenever some vector space $\mathcal{H}$ is a member of $\mathscr{I}$, the successor of $\mathcal{H}$ constructed as\smallskip

\begin{equation}  
   \mathcal{H}^{+}
   \mathrel{\mathop:}=
   \left( \mathcal{H} \cup \{\mathcal{H}\} \right) \ni \mathcal{H}
   \;\;\;\;   
\end{equation}
\smallskip

\noindent is also a member of $\mathscr{I}$.\bigskip

\noindent Be mindful however that the only \emph{singleton vector space} – i.e., a vector space having exactly one element – is $\{0\}$. On top of that, all the successors of $\{0\}$ are one and the same zero vector space $\{0\}$. In fact,\smallskip

\begin{equation}  
   \{0\}^{+}
   \mathrel{\mathop:}=
   \{0\} \cup \left\{\{0\}\right\}
   =
   \{0\} \cup \{0\}
   =
   \{0\}
   \;\;\;\;  ,
\end{equation}
\smallskip

\noindent therefore\smallskip

\begin{equation}  
   \{0\}
   \in
   \{0\}^{+}
   \in
   {\{0\}^{+}}^{+}
   \in
   {{\{0\}^{+}}^{+}}^{+}
   \in
   \dots
   \;\;\;\;  .
\end{equation}
\smallskip

\noindent Hence, the essence of the axiom $\textbf{\texttt{Inf}}_{\mathcal{L}(\texttt{Hil})}$ can be interpreted as follows: The cardinality of a vector space capable of including some vector space $\mathcal{H}$ together with infinitely many successors of $\mathcal{H}$ can be \emph{finite}, in particular 1. To be sure, the propositional sentence\smallskip

\begin{equation} 
   \Phi \left(\mathscr{I}, \mathcal{H} \right)
   =
   \left(\mathcal{H} \in \mathscr{I} \right)
   \sqcap
   \left(\mathcal{H}^{+} \in \mathscr{I} \right)
   \sqcap
   \left({\mathcal{H}^{+}}^{+} \in \mathscr{I} \right)
   \sqcap
   \left({{\mathcal{H}^{+}}^{+}}^{+} \in \mathscr{I} \right)
   \sqcap
   \dots
   \;\;\;\;   
\end{equation}
\smallskip

\noindent can be true if and only if $\mathcal{H}=\{0\}$. So, $\mathrm{card}\left(\mathscr{I}|\Phi \!\left(\mathscr{I}, \{0\} \right)\right)$, the function determining the cardinality of the vector space $\mathscr{I}$, is defined for any $\mathscr{I}$, say, $\{0\}$, for which one finds\smallskip

\begin{equation} 
   \mathrm{card}\Big(\{0\}\Big|\Phi \!\left(\{0\}, \{0\} \right) \!\Big)
   =
   \mathrm{card}\left(\{0\} \right)
   =
   1
   \;\;\;\;  .
\end{equation}
\smallskip

\noindent By contrast, presented in the same way, the axiom $\textbf{\texttt{Inf}}$ states that the cardinality of a set capable of including some set $X$ together with infinitely many successors of $X$ is infinite. To be sure, in the propositional sentence\smallskip

\begin{equation} 
   \Phi \left(\mathbf{I}, X \right)
   =
   \left(X \in \mathbf{I} \right)
   \sqcap
   \left(X^{+} \in \mathbf{I} \right)
   \sqcap
   \left({X^{+}}^{+} \in \mathbf{I} \right)
   \sqcap
   \left({{X^{+}}^{+}}^{+} \in \mathbf{I} \right)
   \sqcap
   \dots
   \;\;\;\;   
\end{equation}
\smallskip

\noindent $X$ is a bound variable; that is, this sentence depends only on $\mathbf{I}$. Therewithal, $\Phi \left(\mathbf{I} \right)$ is true if $\mathbf{I}$ includes infinitely many sets. As a result, $\mathrm{card}\left(\mathbf{I}|\Phi\!\left(\mathbf{I} \right)\right)$ is infinite.\bigskip

\noindent Let us introduce the axiom $\textbf{\texttt{Fin}}$ into set theory, namely:\smallskip

\begin{equation} 
   \textbf{\texttt{Fin}}
   \mathrel{\mathop:}=
   \forall X
      \left(
         X \text{ meets } \textbf{\texttt{C}}_{\,\text{fin}}
      \right)
   \;\;\;\;  ,
\end{equation}
\smallskip

\noindent where $\textbf{\texttt{C}}_{\,\text{fin}}$ is one of the necessary and sufficient conditions of finiteness. Thus, $\textbf{\texttt{C}}_{\,\text{fin}}$ may be\smallskip

\begin{equation} 
   f\!\!: X \to \{1,2,\dots, n \}
   \;\;\;\;  ,
\end{equation}
\smallskip

\noindent where $f$ is a bijection and $n$ is some natural number \cite{Levy}. As it is clear, $\textbf{\texttt{Fin}}$ implies $\neg\textbf{\texttt{Inf}}$. For its part, assuming $\neg\textbf{\texttt{Inf}}$ and $\neg\textbf{\texttt{Empty}}$ one can infer that the cardinality of the set, which is incapable of including some set $X$ and also infinitely many successors of $X$, can be finite, particularly 1. In symbols,\smallskip

\begin{equation}  
   \neg\textbf{\texttt{Inf}}
   \dotplus
   \neg\textbf{\texttt{Empty}}
   \vdash
   \mathrm{card}\Big( \,\mathbf{I} \,\Big| \neg\Phi(\mathbf{I}) \sqcap \left(\varnothing \notin \mathbf{I}\right) \!\Big)
   \;\;\;\;  .
\end{equation}
\smallskip

\noindent Since the predicate $\neg\Phi(\mathbf{I}) \sqcap \left(\varnothing \notin \mathbf{I}\right)$ is true if $\mathbf{I}$ does not include infinitely many sets and $\mathbf{I}$ differs from $\varnothing$, the cardinality of $\mathbf{I}$ is finite and greater than or equal to 1. This last does suggest that\smallskip

\begin{equation}  
   \neg\textbf{\texttt{Inf}}
   \dotplus
   \neg\textbf{\texttt{Empty}}
   \vdash
   \textbf{\texttt{Inf}}_{\mathcal{L}(\texttt{Hil})}
   \;\;\;\;  .
\end{equation}
\smallskip

\noindent In this way, one finds that the axiomatic set theory $\texttt{ST}_{\texttt{Hil}}$ representing Hilbert space theory may be given in the form\smallskip

\begin{equation} 
   \texttt{ST}_{\texttt{Hil}}
   \mathrel{\mathop:}=
   \texttt{ZF}_{\texttt{fin}}
   \;\;\;\;  ,
\end{equation}
\smallskip

\noindent where $\texttt{ZF}_{\texttt{fin}}$ refers to the axioms of the finite set theory\smallskip

\begin{equation} 
  \texttt{ZF}_{\texttt{fin}}
   \mathrel{\mathop:}=
   \left\{
      \textbf{\texttt{Ext}},
      \textbf{\texttt{Spec}},
      \textbf{\texttt{Reg}},
      \textbf{\texttt{PS}},
      \textbf{\texttt{Uni}},
      \textbf{\texttt{Rep}},
      \neg\textbf{\texttt{Inf}},
      \neg\textbf{\texttt{Empty}}
   \right\}
   \;\;\;\;  .
\end{equation}
\smallskip

\noindent As appears, $\texttt{ST}_{\texttt{Hil}}$ and $\texttt{ZF}$, which is presented by\smallskip

\begin{equation} 
  \texttt{ZF}
   \mathrel{\mathop:}=
   \left\{
      \textbf{\texttt{Ext}},
      \textbf{\texttt{Spec}},
      \textbf{\texttt{Reg}},
      \textbf{\texttt{PS}},
      \textbf{\texttt{Uni}},
      \textbf{\texttt{Rep}},
      \textbf{\texttt{Inf}},
      \textbf{\texttt{Empty}}
   \right\}
   \;\;\;\;  ,
\end{equation}
\smallskip

\noindent cannot be in a hierarchical relation: Their formal languages, $\mathcal{L}(\texttt{ZF}_{\texttt{fin}})$ and $\mathcal{L}(\texttt{ZF})$ respectively, are clear distinct, but neither is more expressive than the other. Because of this, the power of Hilbert space theory is incapable of being compared to the power of Zermelo-Fraenkel set theory.\bigskip

\section{The quantum-classical correspondence}  

\noindent Recall that all possible states of a classical mechanical system can be represented such that each corresponds to exactly one point in some multidimensional space $\Gamma$ called \emph{the phase space} of the system \cite{Hirsch}. Therefore, the formal language of classical mechanics, $\mathcal{L}(\texttt{CM})$, can be understood as isomorphic to the formal language of phase space theory, $\mathcal{L}(\texttt{Phase})$, where $\texttt{Phase}$ denotes the collections of axioms of phase space theory. This means that everything that can be expressed through the language $\mathcal{L}(\texttt{CM})$ can also be expressed through the language $\mathcal{L}(\texttt{Phase})$, and vice versa. Symbolically, the isomorphism from $\mathcal{L}(\texttt{CM})$ to $\mathcal{L}(\texttt{Phase})$ can be presented as a bijective function $f\!\!: \mathcal{L}(\texttt{CM}) \to \mathcal{L}(\texttt{Phase})$.\bigskip

\noindent Like a vector space $\mathcal{H}$, a phase space $\Gamma$ is a set with some added features on itself. However, unlike $\mathcal{H}$ for which the added features are fixed (they are linear operations such as addition and scalar multiplication), in case of $\Gamma$ the features are changeable and depend on a system being a subject of study and the questions being considered. Given this, $\texttt{Phase}$ can be seen to be identical to $\texttt{ST}$, the collection of axioms of some set theory, in conjunction with $\textbf{\texttt{X}}_{\texttt{system}}$, the axiom[s] that is [are] determined by a system under study. Using symbols, this can be displayed as the following bijection:\smallskip

\begin{equation} \label{PHAS} 
   f\!\!: \mathcal{L}(\texttt{Phase}) \to \mathcal{L}(\texttt{ST} \dotplus \textbf{\texttt{X}}_{\texttt{system}})
   \;\;\;\;  .
\end{equation}
\smallskip

\noindent On the condition that $\texttt{ST}$ is $\texttt{ZF}$, one gets that in the case of classical mechanics, the T-sentence takes the form\smallskip

\begin{equation}  
   S
   \text{ is true in }
   \texttt{CM}
   \iff
   \Phi_S
   \text{ is true in }
   \mathcal{L}(\texttt{ZF} \dotplus \textbf{\texttt{X}}_{\texttt{system}})
   \;\;\;\;  ,
\end{equation}
\smallskip

\noindent where $S$ denotes some sensible sentence in the object language of $\texttt{CM}$, and $\Phi_S$ is the formula that copies $S$ in the language $\mathcal{L}(\texttt{ZF} \dotplus \textbf{\texttt{X}}_{\texttt{system}})$. For example, let $S$ be ``The value of position in the $x$-direction of a classical mechanical system lies in the interval $X_S$''. Then, $\Phi_S$ is the formula $x \in X_S$ evaluated to true for each given position in the $x$-direction of the system.\bigskip

\noindent Nevertheless, in the belief that quantum mechanics is a fundamental theory and so everything is ultimately describable in quantum-mechanical terms, one must insist that any sentence $S$ in the object language of $\texttt{CM}$ should be true if and only if the copy of $S$ is true in $\mathcal{L}(\texttt{QM})$. This means that the following hierarchy of metalanguages must hold:\smallskip

\begin{equation}  
   S
   \text{ is true in }
   \texttt{CM}
   \iff
   \Phi_S
   \text{ is true in }
   \mathcal{L}(\texttt{ZF} \dotplus \textbf{\texttt{X}}_{\texttt{system}})
   \iff
   \Psi_S
   \text{ is true in }
   \mathcal{L}(\texttt{Hil} \dotplus \textbf{\texttt{X}}_{\texttt{QM}})
   \;\;\;\;  ,
\end{equation}
\smallskip

\noindent where $\Psi_S$ stands for the WFF that copies $S$ in the formalism of $\texttt{QM}$.\bigskip

\noindent The above shows that as long as $f\!\!:\texttt{Hil} \to \texttt{ZF}_{\texttt{fin}}$, i.e., Hilbert space theory can be considered to be synonymous with a theory of sets wherein the axioms of infinity and empty set are negated, the assumption of fundamentality of quantum mechanics requires $\mathcal{L}(\texttt{ZF}_{\texttt{fin}} \dotplus \textbf{\texttt{X}}_{\texttt{QM}})$ to be a metalanguage with respect to $\mathcal{L}(\texttt{ZF} \dotplus \textbf{\texttt{X}}_{\texttt{system}})$:\smallskip

\begin{equation}  
   \phi
   \text{ is true in }
   \mathcal{L}(\texttt{ZF} \dotplus \textbf{\texttt{X}}_{\texttt{system}})
   \iff
   \psi_{\phi}
   \text{ is true in }
   \mathcal{L}(\texttt{ZF}_{\texttt{fin}} \dotplus \textbf{\texttt{X}}_{\texttt{QM}})
   \;\;\;\;  ,
\end{equation}
\smallskip

\noindent where $\phi$ is any WFF in $\mathcal{L}(\texttt{ZF} \dotplus \textbf{\texttt{X}}_{\texttt{system}})$ that may be interpreted as representing a declarative sentence and $\psi_{\phi}$ is its copy in $\mathcal{L}(\texttt{ZF}_{\texttt{fin}} \dotplus \textbf{\texttt{X}}_{\texttt{QM}})$.\bigskip

\noindent But in accordance with what has been demonstrated in the previous section, the formal languages $\mathcal{L}(\texttt{ZF} \dotplus \textbf{\texttt{X}}_{\texttt{system}})$ and $\mathcal{L}(\texttt{ZF}_{\texttt{fin}} \dotplus \textbf{\texttt{X}}_{\texttt{QM}})$ cannot be in a hierarchical relation: Even if $\textbf{\texttt{X}}_{\texttt{system}}$ were to be included in $\textbf{\texttt{X}}_{\texttt{QM}}$, $\texttt{ZF}$ wouldn’t be a part of $\texttt{ZF}_{\texttt{fin}}$. The implication of this fact is that classical mechanics cannot be reducible to quantum mechanics. Such a finding flatly contradicts the usual assumed hierarchy of theories in physics whereby quantum theory should have been more fundamental than its classical counterpart.\bigskip

\section{The emergence of infinities}  

\noindent To get by with non-reducibility of $\texttt{ZF}$ and $\texttt{ZF}_{\texttt{fin}}$, one may assume that there can be no isomorphism from $\texttt{Hil}$ to $\texttt{ZF}_{\texttt{fin}}$; instead, $\texttt{Hil}$ must be isomorphic to $\texttt{ZF}$. In symbols,\smallskip

\begin{equation} \label{CONJ} 
   f\!\!: \texttt{Hil} \to \texttt{ZF}
   \;\;\;\;  .
\end{equation}
\smallskip

\noindent Accordingly,\smallskip

\begin{equation} \label{MAP1} 
   f\!\!: \mathcal{L}(\texttt{CM}) \to \mathcal{L}(\texttt{ZF} \dotplus \textbf{\texttt{X}}_{\texttt{system}})
   \;\;\;\;  ,
\end{equation}

\begin{equation} \label{MAP2} 
   f\!\!: \mathcal{L}(\texttt{QM}) \to \mathcal{L}(\texttt{ZF} \dotplus \textbf{\texttt{X}}_{\texttt{QM}})
   \;\;\;\;  .
\end{equation}
\smallskip

\noindent Subject to the condition $\textbf{\texttt{X}}_{\texttt{QM}} \mathrel{\mathop:}= \textbf{\texttt{X}}_{\texttt{system}} \dotplus \textbf{\texttt{X}}_{\texttt{quantum}}$, where $\textbf{\texttt{X}}_{\texttt{quantum}}$ is [are] some pure quantum axiom[s], $\mathcal{L}(\texttt{QM})$ can be taken to be a metalanguage with respect to $\mathcal{L}(\texttt{CM})$ as proof of fundamentality of quantum mechanics.\bigskip

\noindent The conjecture (\ref{CONJ}) alleges some sort of \emph{a metamathematical imposition}: It assumes that axioms for logic and mathematics must be formulated only on sets endowed with no features at all. Such sets are called \emph{pure} or \emph{hereditary} since all elements of those sets are themselves sets. Contrastively, a vector space – a set with linear operations defined upon it – is not pure: It contains \emph{urelements}, i.e., objects (vectors) that are not sets but may be elements of sets \cite{Weisstein}. So, in accordance with the aforesaid imposition, axioms that are new to or different from the axiomatic system of Zermelo–Fraenkel set theory cannot be acquired by interchanging “sets” with “vector spaces” in the $\texttt{ZF}$ axioms.\bigskip

\noindent But on the other hand, one must admit that there is no deep mathematical reason for preferring pure sets to sets containing urelements \cite{Taylor}. What is even more crucial, the imposition of pure sets (\ref{CONJ}) bring about one of the most serious problems in modern physics, viz., the emergence of infinities in WFFs of the classical and quantum formalisms. Let us elaborate upon this.\bigskip

\noindent On the understanding that the successor of the natural number $n$ is the smallest natural number greater than $n$, specifically,\smallskip

\begin{equation} 
   n^{+}
   \mathrel{\mathop:}=
   n + 1
   \;\;\;\;  ,
\end{equation}
\smallskip

\noindent it follows that\smallskip

\begin{equation} 
   \left(n + 1\right)^{+}
   \mathrel{\mathop:}=
   n + 1 + 1
   \;\;\;\;  ,
\end{equation}

\begin{equation} 
   \left(n + 1 + 1\right)^{+}
   \mathrel{\mathop:}=
   n + 1 + 1 + 1
   \;\;\;\;  ,
\end{equation}
\smallskip

\noindent and so on. As an example, the $N^{\text{th}}$  successor of 0 is the expression of the form $\sum_{n=1}^{N} 1$, i.e., the sum of the terms of the sequence $\left(1\right)_{n=1}^{N}$.\bigskip

\noindent According to the axiom $\textbf{\texttt{Inf}}$ of $\texttt{ZF}$, there is a set $\mathbf{I}$ that includes infinitely many successors of 0. This entails the existence of an infinite sequence of units, namely,\smallskip

\begin{equation} 
   \left(1\right)_{n=1}^{1},
   \left(1\right)_{n=1}^{2},
   \left(1\right)_{n=1}^{3},
   \dots,
   \left(1\right)_{n=1}^{N},
   \ldots
   =
   1,1,1,\dots,1,\ldots
   \;\;\;\;  .
\end{equation}
\smallskip

\noindent Introducing the symbol $\infty$ to denote an unbounded limit, the sum of the terms of this sequence can be represented by the expression\smallskip

\begin{equation} 
   \sum_{n=1}^{\infty}
   1
   =
   + \infty
   \;\;\;\;  ,
\end{equation}
\smallskip

\noindent where $+\infty$ symbolizes \emph{positive infinity} that is added (along with \emph{negative infinity} $-\infty$) to the real number system $\mathbb{R}$ and treated as an actual number. Consequently, the sum of all the successors of 0 must be equal to positive infinity:\smallskip

\begin{equation} 
   \left(0 + 1\right)+
   \left(0 + 1 + 1\right)+
   \left(0 + 1 + 1 + 1\right)+
   \dots
   =
   \sum_{n=1}^{\infty}
   n
   =
   +\infty
   \;\;\;\;  .
\end{equation}
\smallskip

\noindent The notion of unbounded limit provides justification for infinite-dimensional spaces (linear or otherwise) over the fields of real and complex numbers. Furthermore, using the reciprocals of the infinite elements $+\infty$ and $-\infty$, namely, $+\frac{1}{\infty}$ and $-\frac{1}{\infty}$ , one can introduce the concept of \emph{infinitesimals}. In this way, the axiom $\textbf{\texttt{Inf}}$ enables the development of calculus. But in line with that, $\textbf{\texttt{Inf}}$ gives rise to infinities in calculated physical quantities.\bigskip

\noindent Take, for example, the value of the zero-point energy. This value is simply the sum over all the excitation modes of the vacuum, i.e., over all the wavevectors $\mathbf{k}$,

\begin{equation} \label{VAC} 
   \langle 0|\hat{H}_{F}|0 \rangle
   =
   \frac{\hbar}{2}
   \sum_{\mathbf{k}} 
   \left| \omega_{\mathbf{k}} \right|
   \;\;\;\;  ,
\end{equation}
\smallskip

\noindent where $\hat{H}_{F}$ is the particle field Hamiltonian and each $\left| \omega_{\mathbf{k}} \right|$ is the modulus of the energy mode. Now consider a box of volume $L^3$ and let the wavelength along, say, the $x$-axis be $\lambda_x = \frac{L}{n_x}$ for some natural number $n_x$. Then, in the interval $\left( k_x, k_x + \Delta k_x \right)$, there would be $\frac{\Delta k_x L}{2\pi} = \Delta n_x$ discrete values of $k_x$. Consequently, the expression (\ref{VAC}) would become\smallskip

\begin{equation} 
   \langle 0|\hat{H}_{F}|0 \rangle
   =
   \frac{\hbar L^3}{2}
   \sum_{\mathbf{k}}
   \frac{\left( \Delta \mathbf{k} \right)^3}{\left( 2\pi \right)^3}
   \left| \omega_{\mathbf{k}} \right|
   \;\;\;\;  ,
\end{equation}
\smallskip

\noindent where $\left( \Delta \mathbf{k} \right)^3$ denotes the product $\Delta k_x \Delta k_y \Delta k_z$ and $\left| \omega_{\mathbf{k}} \right|$ is\smallskip

\begin{equation} 
   \left| \omega_{\mathbf{k}} \right|
   =
   c
   \,\sqrt{
      \left( \frac{m_0 c}{\hbar} \right)^2
      +
       k_{x}^2 + k_{y}^2 + k_{z}^2
   }
   \;\;\;\;  .
\end{equation}
\smallskip

\noindent For the sake of simplification, assume that particles are massless (i.e., $m_0 = 0$), $k_{x}=k_{y}=k_{z}=\frac{2\pi}{L}n$, and $\Delta n=1$. After that,\smallskip

\begin{equation}  
   \langle 0|\hat{H}_{F}|0 \rangle
   =
   \frac{\sqrt{3} \pi \hbar c}{L}
   \sum_{n=1} 
   n
   \;\;\;\;  .
\end{equation}
\smallskip

\noindent 

\noindent Since the axiom $\textbf{\texttt{Inf}}$ is a part of the quantum formalism, the existence of an endless sequence of the natural numbers $\left( n \right)_{n=1}^{\infty}$ must be allowed therein. The implication of this is that the vacuum energy in any finite volume of space ends up being determined by the divergent series\smallskip

\begin{equation}  
   \langle 0|\hat{H}_{F}|0 \rangle
   =
   \frac{\sqrt{3} \pi \hbar c}{L}
   \sum_{n=1}^{\infty} 
   n
   =
   +\infty
   \;\;\;\;  .
\end{equation}
\smallskip

\noindent To circumvent this result, one may argue that only differences in energy are physically measurable, while the total vacuum energy is not observable. So, to obtain the Hamiltonian with observable energy, one may subtract the vacuum expectation value $\langle 0|\hat{H}_{F}|0 \rangle$ from the Hamiltonian $\hat{H}_{F}$. This line of reasoning is the basis of renormalization.\bigskip

\noindent That said, when dealing with extended real numbers $\mathbb{R}\cup\{-\infty,+\infty\}$, the expression $\infty - \infty$ is left undefined.\bigskip

\noindent So, to make the act of subtracting $\langle 0|\hat{H}_{F}|0 \rangle$ from $\hat{H}_{F}$ well-defined, one needs a regularization process (which is the first step of any renormalization). To this end, the technique known as \emph{analytic continuation} can be used. In a little more detail, the Riemann zeta function $\zeta$ of a complex variable $s$ is defined in the region $\mathrm{Re}[s]>1$ by the absolutely convergent series\smallskip

\begin{equation}  
   \zeta(s)
   =
   \sum_{n=1}^{\infty} 
   n^{-s}
   \;\;\;\;  .
\end{equation}
\smallskip

\noindent In the region $\mathrm{Re}[s]\le 1$, though, this series is convergent neither absolutely nor conditionally. Still, the function $\zeta$ can be extended in that region by analytic continuation \cite{Elizalde}. For instance, it can be shown that the following formula holds true:\smallskip

\begin{equation}  
   \zeta(-s)
   =
   \sum_{n=1}^{\infty} 
   n^{s}
   =
   -\frac{B_{s+1}}{s+1}
   \;\;\;\;  ,
\end{equation}
\smallskip

\noindent where $B_{s+1}$ are the Bernoulli numbers (e.g., $B_1=\frac{1}{2}$ and $B_2=\frac{1}{6}$). The above provides a pretext for the following regularized sums:\smallskip

\begin{equation}  
   \sum_{n=1}^{\infty} 
   1
   =
   -\frac{1}{2}
   \;\;\;\;  ,
\end{equation}

\begin{equation}  
   \sum_{n=1}^{\infty} 
   n
   =
   -\frac{1}{12}
   \;\;\;\;  .
\end{equation}
\smallskip

\noindent Accordingly, the value $\langle 0|\hat{H}_{F}|0 \rangle$ turns into finite and the formula for observable energy\smallskip

\begin{equation}  
   \hat{H}_{F\,\text{observable}}
   =
   \hat{H}_{F}
   -
   \langle 0|\hat{H}_{F}|0 \rangle
   \;\;\;\;   
\end{equation}
\smallskip

\noindent becomes well-defined.\bigskip

\noindent Nonetheless, the question remains: If calculations of physical quantities are assumed from the axiom of infinity $\textbf{\texttt{Inf}}$ of $\texttt{ZF}$, can always be possible to eliminate infinities occurring in those calculations?\bigskip

\noindent The next section will show in the most general terms that the answer to this question must be a ``no''.\bigskip

\section{Quantizability}  

\noindent Let a classical field theory $\texttt{T}$ be completely described by a certain function $Z$ of the state variables $x_i$ and coupling constants $J_k$. By way of illustration, let $Z$ be a Hamiltonian $T+V$ whose kinetic and interaction parts, i.e., $T$ and $V$ respectively, depend on $x_i$. Then, the coupling constant $J$ will be the magnitude of the $T$ part with respect to the $V$ part. Since $Z$ constitutes the whole description of $\texttt{T}$, this function must determine $\Sigma_{\Phi}$, the set of all well-defined formulae $\Phi$ of $\mathcal{L}(\texttt{T})$.\bigskip

\noindent Assume that the theory $\texttt{T}$ can be quantized by a linear map $\{x_i\} \to \{\tilde{x}_i\}$ and a certain change in the couplings $\{J_k\} \to \{\tilde{J}_k\}$ such that the function $Z$ can be rewritten in terms of the quantum variables $\tilde{x}_i$ only. In that case, the set $\Sigma_{\Phi}$ becomes converted into $\Sigma_{\tilde{\Phi}}$, the set of formulae $\tilde{\Phi}$ belonging to the formalism $\mathcal{L}(\tilde{\texttt{T}})$ of the quantized field theory $\tilde{\texttt{T}}$.\bigskip

\noindent Therewith, in conformity with reductionism, the axiomatization of $\tilde{\texttt{T}}$ (i.e., the set of axioms of the quantized field theory depicted by $\{\tilde{\texttt{T}}\}$) must be presented by the collection of axioms of the classical field theory (entitled $\{\texttt{T}\}$) along with the axioms of $\tilde{\texttt{T}}$ absent in $\texttt{T}$ (denoted by $\textbf{\texttt{X}}_{\tilde{\texttt{T}}})$. This suggests the isomorphism\smallskip

\begin{equation}  
   f\!\!: \{\tilde{\texttt{T}}\} \to \{\texttt{T}\} \dotplus \textbf{\texttt{X}}_{\tilde{\texttt{T}}}
   \;\;\;\;   .
\end{equation}
\smallskip

\noindent If all formulae $\tilde{\Phi}$ are free from infinities, then $\texttt{T}$ is called \emph{quantizable} \cite{Kaku}. In symbols, the requirement of quantizability can be presented in the form of the material biconditional:\smallskip

\begin{equation}  
   \texttt{T}
   \text{ is quantizable}
   \iff
   Q
   \;\;\;\;  ,
\end{equation}
\smallskip

\noindent where $Q$ is the statement\smallskip

\begin{equation}  
   Q
   \mathrel{\mathop:}=
   \forall \Phi \in \mathcal{L}(\{\texttt{T}\})
      \left[
         \exists \tilde{\Phi} \in \mathcal{L}(\{\texttt{T}\} \dotplus \textbf{\texttt{X}}_{\tilde{\texttt{T}}})
         \left(
            \tilde{\Phi} \not \owns \pm\infty
         \right)
      \right]
   \;\;\;\;  ,
\end{equation}
\smallskip

\noindent which is true if for each well-defined classical formula $\Phi$ there exists a well-defined formula $\tilde{\Phi}$ of the quantized formalism such that $\tilde{\Phi}$ does not contain the infinite elements $+\infty$ and $-\infty$. If $\texttt{T}$ is quantizable and, on top of that, $\{\tilde{J}_k\}$ – the set of the couplings that have been fixed – is finite, $\texttt{T}$ is called \emph{renormalizable}, otherwise $\texttt{T}$ is designated as a quantizable nonrenormalizable classical field theory \cite{Kaku}.\bigskip

\noindent Let $\texttt{T}$ be such that its axioms $\{\texttt{T}\}$ match up the $\texttt{ZF}$ axioms coupled with additional axioms $\textbf{\texttt{X}}_{\texttt{field}}$ depending on a classical filed under study, i.e.,\smallskip

\begin{equation}  
   f\!\!: \{\texttt{T}\} \to \texttt{ZF} \dotplus \textbf{\texttt{X}}_{\texttt{field}}
   \;\;\;\;   .
\end{equation}
\smallskip

\noindent Allowing that $\textbf{\texttt{X}}_{\tilde{\texttt{T}}} \mathrel{\mathop:}= \textbf{\texttt{X}}_{\texttt{field}} \dotplus \textbf{\texttt{X}}_{\texttt{quantum}}$, the statement $Q$ then becomes the proposition\smallskip

\begin{equation}  
   Q
   \mathrel{\mathop:}=
   \forall \Phi \in \mathcal{L}(\texttt{ZF} \dotplus \textbf{\texttt{X}}_{\texttt{field}})
      \left[
         \exists \tilde{\Phi} \in \mathcal{L}(\texttt{ZF} \dotplus \textbf{\texttt{X}}_{\texttt{field}} \dotplus \textbf{\texttt{X}}_{\texttt{quantum}})
         \left(
            \tilde{\Phi} \not \owns \pm\infty
         \right)
      \right]
   \;\;\;\;   
\end{equation}
\smallskip

\noindent whose predicate ``$\not \owns \pm\infty$'' is not contained in its subject ``$\tilde{\Phi} \in \mathcal{L}(\texttt{ZF} \dotplus \textbf{\texttt{X}}_{\texttt{field}} \dotplus \textbf{\texttt{X}}_{\texttt{quantum}})$''. By virtue of universal instantiation, it may be inferred that the property of being free from infinites cannot hold for all formalisms embracing the axiom $\textbf{\texttt{Inf}}$. As a result, the statement $Q$ is logically contingent on $\textbf{\texttt{X}}_{\texttt{field}}$ and $\textbf{\texttt{X}}_{\texttt{quantum}}$ (i.e., the specific classical and quantum axiomatizations). Ergo, there is no way to rule out an occasion when $Q$ gets to be false and so infinities cannot be avoided in $\tilde{\Phi}$.\bigskip

\noindent By contrast, let us suppose that the axiom $\neg\textbf{\texttt{Inf}}$ of the finite set theory $\texttt{ZF}_{\texttt{fin}}$ is a necessary part of the formalism of a theory $\texttt{T}$. In which case, the statement $Q$ turns into the proposition\smallskip

\begin{equation}  
   Q
   \mathrel{\mathop:}=
   \forall \Phi \in \mathcal{L}(\texttt{ZF}_{\texttt{fin}} \dotplus \textbf{\texttt{X}}_{\texttt{field}})
      \left[
         \exists \tilde{\Phi} \in \mathcal{L}(\texttt{ZF}_{\texttt{fin}} \dotplus \textbf{\texttt{X}}_{\texttt{field}} \dotplus \textbf{\texttt{X}}_{\texttt{quantum}})
         \left(
            \tilde{\Phi} \not \owns \pm\infty
         \right)
      \right]
   \;\;\;\;   
\end{equation}
\smallskip

\noindent whose predicate ``$\not \owns \pm\infty$'' is embodied in the subject ``$\tilde{\Phi} \in \mathcal{L}(\texttt{ZF}_{\texttt{fin}} \dotplus \textbf{\texttt{X}}_{\texttt{field}} \dotplus \textbf{\texttt{X}}_{\texttt{quantum}})$''. On account of universal generalization, this means that the property of being free from infinites must hold for all formalisms embracing the axiom of finiteness $\neg\textbf{\texttt{Inf}}$. Consequently, the statement $Q$ becomes a strict logical truth: It can be validated by the specific classical and quantum axiomatizations but is not grounded in them.\bigskip

\noindent This entails the following conclusion: Motivated by reductionism, a belief that any classical field theory – including general relativity – is quantizable, cannot hold true within the limits of Zermelo-Fraenkel set theory. On the contrary, against the backdrop of $\texttt{ZF}_{\texttt{fin}}$, it is ensured that all classical fields are quantizable (in full agreement with reductionism).\bigskip

\section{In the absence of the axiom of infinity}  

\noindent One may suppose that infinities are inevitable part of nature and the axiom $\textbf{\texttt{Inf}}$, which essentially states that an infinite procedure can be considered as a single element, only reflects such fact. This supposition entails the existence of at least one classical field theory that is not quantizable along with the existence of quantizable but nonrenormalizable classical field theories.\bigskip

\noindent Besides, given that certain functions cannot be extended in a Taylor series at some number[s], the decomposition of solutions to the classical equations of motion into sums of normal (i.e., orthogonal) modes may not be used to quantize field theories that are described by such functions. To reconcile those theories with the principles of quantum mechanics, approaches different from perturbative theory should be followed. Among them, string theory and loop gravity are the most popular.\bigskip

\noindent Be that as it may, the question of non-perturbative quantum field theory is a formidable one. To begin with, non-perturbative solutions to the classical equations of motion are generally hard or even intractable. Further, when one uses a non-perturbative quantization of general relativity, it is difficult to ensure that one recovers real general relativity as opposed to complex general relativity \cite{Nicolai}. Not just that, it is believed by many that any realistic interaction should be interpreted as a sum of intermediate states involving the exchange of various virtual particles. However, such an interpretation makes sense only in the framework of perturbation theory.\bigskip

\noindent From here it can be inferred that non-reducibility of $\texttt{ZF}$ and $\texttt{ZF}_{\texttt{fin}}$ must be treated in a different way. Namely, it must be assumed that in the bijection $f:\mathcal{L}(\texttt{Phase}) \to \mathcal{L}(\texttt{ST} \dotplus \textbf{\texttt{X}}_{\texttt{system}})$, $\texttt{ST}$ is $\texttt{ZF}_{\texttt{fin}}$, not $\texttt{ZF}$.\bigskip

\noindent In view of this, the formal languages of classical and quantum mechanics are such that\smallskip

\begin{equation} \label{MAP3} 
   f\!\!: \mathcal{L}(\texttt{CM}) \to \mathcal{L}(\texttt{ZF}_{\texttt{fin}} \dotplus \textbf{\texttt{X}}_{\texttt{system}})
   \;\;\;\;  ,
\end{equation}

\begin{equation} \label{MAP4} 
   f\!\!: \mathcal{L}(\texttt{QM}) \to \mathcal{L}(\texttt{ZF}_{\texttt{fin}} \dotplus \textbf{\texttt{X}}_{\texttt{system}} \dotplus \textbf{\texttt{X}}_{\texttt{quantum}})
   \;\;\;\;  .
\end{equation}
\smallskip

\noindent Consistent with the above mappings, a phase space $\Gamma$ of a classical mechanical system has a finite number of points. Likewise, a Hilbert space $\mathcal{H}$ of a quantum mechanical system also needs to be a finite set of vectors over a finite field $\mathbb{F}$ that contains $p^k$ elements (where $p$ is a prime number and $k$ is a positive integer). Accordingly, the cardinality of $\mathcal{H}$ is finite and defined by\smallskip

\begin{equation}  
   \mathrm{card}\left( \mathcal{H} \right)
   =
   p^{k \,\mathrm{dim}(\mathcal{H})}
   \;\;\;\;  ,
\end{equation}
\smallskip

\noindent where $\mathrm{dim}(\mathcal{H})$ is the dimension of $\mathcal{H}$.\bigskip

\noindent The following square summable sequence can be taken as an example of such a Hilbert space:\smallskip

\begin{equation}  
   l_{2}
   =
   \left\{
      c_{n} \in R_p
      \left|
      \,\sum_{n=0}^{N}|c_{n}|^{2}
      \right.
   \right\}
   \;\;\;\;  ,
\end{equation}
\smallskip

\noindent where $R_p$ is a set of $p^2$ elements of the finite field $\mathbb{E}_p$ containing $p$ elements, namely,\smallskip

\begin{equation}  
   R_p
   =
   \left\{
      x+iy
      \left|
      \,x,y \in \mathbb{E}_p
      \right.
   \right\}
   \;\;\;\;  .
\end{equation}
\smallskip

\noindent This set is endowed with multiplicative inverses as well addition and multiplication defined by\smallskip

\begin{equation}  
   (x+iy)
   +
   (z+iw)
   \mathrel{\mathop:}=
   (x+z)
   +
   i(y+w)
   \;\;\;\;  ,
\end{equation}

\begin{equation}  
   (x+iy)
   (z+iw)
   \mathrel{\mathop:}=
   (xz-yw)
   +
   i(xw+yz)
   \;\;\;\;  .
\end{equation}
\smallskip

\noindent As to subtraction and division (excluding division by zero), they are inherited by $R_p$ from the field $\mathbb{E}_p$. All this implies that $R_p$ is a finite field.\bigskip

\noindent In line with the mappings (\ref{MAP3}) and (\ref{MAP4}), the conventional hierarchy of physical theories (wherein each quantum field theory $\tilde{\texttt{T}}$ is more fundamental than its classical counterpart $\texttt{T}$) holds because\smallskip

\begin{equation}  
   \mathrm{card}\left(\Sigma_{\texttt{T}}\right) < \mathrm{card}\left(\Sigma_{\tilde{\texttt{T}}}\right)
   \;\;\;\;  ,
\end{equation}
\smallskip

\noindent where $\Sigma_{\texttt{T}}$ and $\Sigma_{\tilde{\texttt{T}}}$ are the sets of formulae of $\mathcal{L}(\texttt{ZF}_{\texttt{fin}} \dotplus \textbf{\texttt{X}}_{\texttt{field}})$ and $\mathcal{L}(\texttt{ZF}_{\texttt{fin}} \dotplus \textbf{\texttt{X}}_{\texttt{field}} \dotplus \textbf{\texttt{X}}_{\texttt{quantum}})$ respectively.\bigskip

\noindent More importantly, provided that any quantum field theory $\tilde{\texttt{T}}$ is decidable, every WFF in $\mathcal{L}(\texttt{ZF}_{\texttt{fin}} \dotplus \textbf{\texttt{X}}_{\texttt{field}} \dotplus \textbf{\texttt{X}}_{\texttt{quantum}})$ must be consistent with the axioms $\texttt{ZF}_{\texttt{fin}}$. This implies that calculations made with $\tilde{\texttt{T}}$ are expected to be free from the infinite elements $+\infty$ and $-\infty$.\bigskip

\noindent To be sure, let us go back to the calculation of the vacuum expectation value of the energy of a particle field. The result of summation over the first $N$ excitation modes of the vacuum depends on the natural number $N$ as follows:\smallskip

\begin{equation}  
   {\langle 0|\hat{H}_{F}|0 \rangle}_{\! N}
   =
   \frac{\sqrt{3} \pi \hbar c}{L}
   \sum_{n=1}^{N} 
   n
   =
   \frac{\sqrt{3} \pi \hbar c}{L}
   \frac{N(N+1)}{2}
   \;\;\;\;  .
\end{equation}
\smallskip

\noindent Now recall that the axiom $\neg\textbf{\texttt{Inf}}$ of the finite set theory $\texttt{ZF}_{\texttt{fin}}$ negates the existence of a set that includes all the natural numbers. This indicates that the quantification over the infinite domain of discourse $\mathbb{N}$ consisting of all the natural numbers, that is,\smallskip

\begin{equation}  
   \forall N \in \mathbb{N}
   \left(
      \!{\langle 0|\hat{H}_{F}|0 \rangle}_{\! N}
      =
      \frac{\sqrt{3} \pi \hbar c}{L}
      \frac{N(N+1)}{2}
   \right)
   \;\;\;\;   ,
\end{equation}
\smallskip

\noindent is nonsensical. But so is the infinite sequence $\left({\langle 0|\hat{H}_{F}|0 \rangle}_{\! N}\right)_{N=1}^{\infty}$.\bigskip

\noindent On the other hand, since the set $S$ containing the partial sums ${\langle 0|T_{00}|0 \rangle}_{\! N}$ is finite, there exists a bijection\smallskip

\begin{equation}  
   f\!\!: S \to
   \left\{ {\langle 0|T_{00}|0 \rangle}_{\!N} \left| \, N < K \right. \right\}
   \;\;\;\;    
\end{equation}
\smallskip

\noindent for some natural number $K$. Consequently, the zero-point energy of the vacuum in a box of volume $L^3$ is finite and can be evaluated by the expression:\smallskip

\begin{equation} \label{TZPE} 
   \langle 0|\hat{H}_{F}|0 \rangle
   <
   \frac{\sqrt{3} \pi \hbar c}{L}
   \frac{K(K+1)}{2}
   \;\;\;\;  .
\end{equation}
\smallskip

\noindent In order to meet the demand on every WFF in a quantum formalism to be consistent with the axioms of the finite set theory $\texttt{ZF}_{\texttt{fin}}$, empty space must have a finite number of points.\bigskip

\noindent Indeed, the calculation of the vacuum energy $\langle 0|\hat{H}_{F}|0 \rangle$ can be viewed as a summation over all quantum harmonic oscillators with the zero-point energy $\frac{\hbar \omega}{2}$ at all $P(L)$ points in space enclosed by an empty cube of edge length $L$:\smallskip

\begin{equation}  
   \langle 0|\hat{H}_{F}|0 \rangle
   =
   \frac{\hbar \omega}{2}
   \sum_{n=1}^{P(L)} 1
   =
   \frac{\hbar \omega}{2}
   P(L)
   \;\;\;\;  .
\end{equation}
\smallskip

\noindent Granted that $\omega =\frac{2\pi c}{L}$, one finds that this summation must bring a finite quantity in compliance with (\ref{TZPE}):\smallskip

\begin{equation}  
   P(L)
   <
   \frac{\sqrt{3}}{2}
   K(K+1)
   \;\;\;\;  .
\end{equation}
\smallskip

\noindent However, such could be allowed to happen only if empty space were to have no more than a finite number of points.\bigskip

\noindent To put it in more dramatic terms, the energy of the vacuum can be held as a measure of the number of points that make up empty space. By way of illustration, let us estimate how many points are contained in empty space of the observable universe. Considering that $\pi \hbar c \,P\!\left( L_{_\mathrm{U}}\!\right) \!/ L_{_\mathrm{U}}$, the vacuum energy in an empty cube of edge length equal to the diameter of the observable universe $L_{_\mathrm{U}}$, is $\rho_{\mathrm{vac}} \cdot L_{_\mathrm{U}}^3$, where $\rho_{\mathrm{vac}}$ is the vacuum energy density, one can write the equality\smallskip

\begin{equation} \label{EQUL} 
   P\!\left( L_{_\mathrm{U}}\!\right)
   =
   \frac{ \rho_{\mathrm{vac}} \cdot L_{_\mathrm{U}}^4 }
  {\pi \hbar c}
   \;\;\;\;  .
\end{equation}
\smallskip

\noindent Based on cosmological observations \cite{Planck}, $\rho_{\mathrm{vac}}$ is $5.4 \times 10^{-10} \,\mathrm{J} \cdot \mathrm{m}^{-3}$ and $L_{_\mathrm{U}}$ is $8.8 \times 10^{26} \,\mathrm{m}$; hence, it can be concluded that the vacuum in the observable universe has at most $3.3 \times 10^{123}$ points.\bigskip

\noindent On the other hand, so long as free space contains a finite number of points, it gives reason to believe that a geometry of the observable universe is finite.\bigskip

\section{A universe with a finite geometry}  

\noindent Let us make a few remarks about the physical universe whose geometry is finite.\bigskip

\noindent But before that, allow us to clarify the difference between finite and discrete geometries. For the purpose of the current presentation, one can define a geometry as a system of axioms that identify what “things” are which constitute “points”, “lines”, “planes”, and so forth. In terms of this definition, a finite geometry is any of axiomatic systems that have only a finite number of points.\bigskip

\noindent By contrast, a discrete geometry (including the causal set program \cite{Brightwell, Sorkin} whereby spacetime is a collection of discrete points) takes up only objects in which points are isolated from each other in some sense. As for example, the set of natural numbers $\mathbb{N}$ is a discrete set, i.e., a set of isolated points. Most important of all, a discrete object (e.g., the set $\mathbb{N}$ or the causal set representing discrete spacetime events) need not have a finite number of points \cite{Bezdek}.\bigskip

\subsection{General relativity in a finite geometry} 

\noindent Denote a geometry over a field $\mathbb{F}$ by $\mathscr{A}$. Consider a set of points (\emph{pointset} for short) $\mathcal{P}_{\mathscr{V}}$ in $\mathscr{A}$ and its proper subset $\mathcal{P}_{\mathscr{U}} \subset \mathcal{P}_{\mathscr{V}}$. Their cardinalities are\smallskip

\begin{equation}  
   \mathrm{card}\left( \mathcal{P}_{\mathscr{V}}\right)
   =
   \mathrm{card}\left( \mathbb{F}^{\prime} \right)^{\mathrm{dim}(\mathscr{A})}
   \;\;\;\;  ,
\end{equation}

\begin{equation}  
   \mathrm{card}\left( \mathcal{P}_{\mathscr{U}}\right)
   =
   \mathrm{card}\left( \mathbb{F}^{\prime\prime} \right)^{\mathrm{dim}(\mathscr{A})}
   \;\;\;\;  ,
\end{equation}
\smallskip

\noindent where $\mathbb{F}^{\prime}$ and $\mathbb{F}^{\prime\prime}$ are subfields of $\mathbb{F}$, i.e., the subsets $\mathbb{F}^{\prime\prime} \subset \mathbb{F}^{\prime} \subseteq \mathbb{F}$ that are fields with respect to the field operations inherited from $\mathbb{F}$, and $\mathrm{dim}(\mathscr{A})$ is the dimension of $\mathscr{A}$. If $\mathcal{P}_{\mathscr{U}}$ has fewer points than $\mathcal{P}_{\mathscr{V}}$, then $\mathscr{A}$ is finite (to be exact, Dedekind-finite), otherwise $\mathscr{A}$ is infinite. In symbols,\smallskip

\begin{equation}  
   \forall \mathcal{P}_{\mathscr{V}} \forall \mathcal{P}_{\mathscr{U}}
   \Big[
      \mathcal{P}_{\mathscr{V}} \!\subseteq\! \mathcal{P}_{\!\mathscr{A}}
      \sqcap
      \mathcal{P}_{\mathscr{U}} \!\subset\! \mathcal{P}_{\!\mathscr{A}}
      \sqcap
      \mathcal{P}_{\mathscr{U}} \!\subset\! \mathcal{P}_{\mathscr{V}}
      \to
      \mathrm{card}\left( \mathcal{P}_{\mathscr{U}}\right) \!<\! \mathrm{card}\left( \mathcal{P}_{\!\mathscr{V}}\right)
   \Big]
   \iff
   \mathcal{P}_{\!\mathscr{A}} \text{ is finite}
   \;\;\;\;  ,
\end{equation}
\vspace{-26pt}

\begin{equation}  
   \forall \mathcal{P}_{\mathscr{V}} \forall \mathcal{P}_{\mathscr{U}}
   \Big[
      \mathcal{P}_{\mathscr{V}} \!\subseteq\! \mathcal{P}_{\!\mathscr{A}}
      \sqcap
      \mathcal{P}_{\mathscr{U}} \!\subset\! \mathcal{P}_{\!\mathscr{A}}
      \sqcap
      \mathcal{P}_{\mathscr{U}} \!\subset\! \mathcal{P}_{\mathscr{V}}
      \to
      \mathrm{card}\left( \mathcal{P}_{\mathscr{U}}\right) \!=\! \mathrm{card}\left( \mathcal{P}_{\!\mathscr{V}}\right)
   \Big]
   \iff
   \mathcal{P}_{\!\mathscr{A}} \text{ is infinite}
   \;\;\;\;  ,
\end{equation}
\smallskip

\noindent where $\mathcal{P}_{\!\mathscr{A}}$ is the pointset of $\mathscr{A}$ itself.\bigskip

\noindent Suppose that $\mathcal{P}_{\mathscr{V}}$ and $\mathcal{P}_{\!\mathscr{U}}$ are enormously large (for the sake of example, $\mathrm{card}\left( \mathcal{P}_{\mathscr{U}}\right)$ and $\mathrm{card}\left( \mathcal{P}_{\mathscr{U}}\right)$ differ from the number $3.3 \times 10^{123}$ by few orders of magnitude at most). Obviously, in that case it is not feasible to compare them by their size. But without practical means to compare the cardinalities of a pointset and its proper subset, one cannot distinguish between finite and infinite geometries. Accordingly, on condition that computational resources are limited, a finite geometry whose pointset is vast can be regarded as being similar to an infinite geometry.\bigskip

\noindent To be specific, in such circumstances, a distance function (a metric) $d$ on a finite geometry $\mathscr{A}$ can be a mapping $\mathcal{P}_{\!\mathscr{A}} \times \mathcal{P}_{\!\mathscr{A}} \to \mathbb{R}$ satisfying all the following metric axioms:\smallskip

\begin{description}[itemindent=+0.2cm]
\item[$\mathbf{(M1)}$] $\forall x,y \in \mathcal{P}_{\!\mathscr{A}} \,\Big( d(x,y) \ge 0 \Big),$
\item[$\mathbf{(M2)}$] $\forall x,y \in \mathcal{P}_{\!\mathscr{A}} \,\Big( d(x,y) = d(y,x) \Big),$
\item[$\mathbf{(M3)}$] $\forall x,y \in \mathcal{P}_{\!\mathscr{A}} \,\Big( x \neq y \,\to\, d(x,y) \neq 0 \Big),$
\item[$\mathbf{(M4)}$] $\forall x,y,z \in \mathcal{P}_{\!\mathscr{A}} \,\Big( d(x,z) \le d(x,y) + d(y,z) \Big).$
\end{description}

\noindent The use of infinite distances may be thought of just as distances greater than others in a system. Consequently, in formulae for the distance functions, one can replace some element of the vast finite field $\mathbb{F}$ by $\infty$ provided that $\infty \cdot 0 = 0$ and $\infty +a= \infty$ for any $a\in\mathbb{F}$ (otherwise, instead of $\infty$, one can use any number larger than the others in the system; however, doing so tends to lead to clumsy proofs \cite{Evans}).\bigskip

\noindent Taking distances as real numbers (rather than elements of finite fields) allows one to apply calculus. The last means that in case of vast pointsets, the field equations of general relativity may be applicable to a finite geometry.\bigskip

\noindent Recall that the possibility of a positive value for the cosmological constant $\Lambda$ present in those equations is implied by the fact that the expansion of the observable universe is accelerating \cite{Hobson}. What is more, the positive $\Lambda$ can be interpreted as the vacuum energy which arises in quantum mechanics \cite{Rugh}. Accordingly, one is allowed to say that $\Lambda = 8\pi G c^{-4}\rho_{\mathrm{vac}}$, or\smallskip

\begin{equation} \label{LAMB} 
   \Lambda
   =
   \frac{8 \pi^{2} \hbar G}{c^{3}}
   \cdot
   \frac{P\!\left( L_{_\mathrm{U}}\right)}{L_{_\mathrm{U}}^4}
   \;\;\;\;  .
\end{equation}
\smallskip

\noindent Let ${L_{_\mathrm{U}}}_{0}$ denote the value of $L_{_\mathrm{U}}$ at the present time $t_0$; then, the diameter of the observable universe in any moment $t$ from now (so that $t - t_{0} = \Delta t \ge 0$) can be written in accordance with Hubble's law as follows\smallskip:

\begin{equation} \label{HUBL} 
   L_{_\mathrm{U}}\!\left( \Delta t \right)
   =
   {L_{_\mathrm{U}}}_{0}
   +
   H
   {L_{_\mathrm{U}}}_{0}
   \Delta t
   \;\;\;\;  ,
\end{equation}
\smallskip

\noindent where $H$ is the time-dependent Hubble parameter which can be expressed in terms of the expansion (or scale factor) $a(\Delta t)$ determining the relative changes in the distance between two gravitationally unbounded objects in the universe during the interval $\Delta t$, namely, $H = {\dot{a}}/{a}$.\bigskip

\noindent Replacing $L_{_\mathrm{U}}$ in (\ref{LAMB}) by (\ref{HUBL}) and providing $H \Delta t$ is small earns the formula for the dependence of the cardinality of the vacuum (i.e., the number of points that constitute empty space) on the interval $\Delta t$:\smallskip

\begin{equation} \label{EXP1} 
   P(\Delta t)
   =
   \frac{c^3 \Lambda}
   {8 \pi^{2} \hbar G}
   \cdot
   {L_{{_\mathrm{U}}}^4}_{0}
   \left(
      1 + 4 \,\frac{\dot{a}}{a} \,\Delta t
   \right)
   \;\;\;\;  .
\end{equation}
\smallskip

\noindent With the provision that the observable universe is expanding, i.e., $L_{_\mathrm{U}}\!\left( \Delta t \right) > {L_{_\mathrm{U}}}_{0}$, it is true to say that $P(\Delta t) > P_{0}$, where $P_{0}$ stands for the current value of $P( \Delta t)$, i.e., $P_{0} = P(0)$. In that case, the first derivative of $P(\Delta t)$ with respect to time $t$ need to be positive.\bigskip

\noindent On the other hand, it follows from (\ref{EXP1}) that this derivative is\smallskip

\begin{equation} \label{DER1} 
   \dot{P}
   =
   \frac{c^3 \Lambda}
   {2 \pi^{2} \hbar G}
   \cdot
   {L_{{_\mathrm{U}}}^4}_{0}
   \left[
      \left(
         \frac{\ddot{a}}{a}
         -
         \left(
            \frac{\dot{a}}{a}
         \right)^2
      \right)\!\Delta t
      +
      \frac{\dot{a}}{a}
   \right]
   \;\;\;\;   .
\end{equation}
\smallskip

\noindent In general relativity, the acceleration equation describing the evolution of the scale factor over time satisfies\smallskip

\begin{equation} \label{ACCE} 
   \frac{\ddot{a}}{a}
   =
   -
   \frac{4 \pi G}{3}
   \cdot
   \left(
      \rho
      +
      \frac{3p}{c^2}
   \right)
   +
   \frac{\Lambda c^2}{a^2}
   \;\;\;\;  ,
\end{equation}
\smallskip

\noindent where $\rho(\Delta t)$ and $p(\Delta t)$ are the mass density and pressure, respectively, that meet the local energy conservation law:\smallskip

\begin{equation} \label{RVAC} 
   \dot{\rho}
   =
   -
   3 \,\frac{\dot{a}}{a}
   \cdot
   \left(
      \rho
      +
      \frac{p}{c^2}
   \right)
   \;\;\;\;  .
\end{equation}
\smallskip

\noindent The first integral of (\ref{ACCE}) and (\ref{RVAC}) is the Friedmann equation\smallskip

\begin{equation}  
   \left(
      \frac{\dot{a}}{a}
   \right)^2
   =
   \frac{8 \pi G}{3} \rho
   -
   \frac{\kappa c^2}{a^2}
   +
   \frac{\Lambda c^2}{a^2}
   \;\;\;\;  ,
\end{equation}
\smallskip

\noindent where $\frac{\kappa}{a^2}$ denotes the spatial curvature of the universe ($\dim \frac{\kappa}{a^2} =\texttt{L}^{-2}$); $\kappa$ belongs to the set $\{-1,0,1\}$ (for negative, zero, and positive curvature respectively), and so $a(\Delta t)$ has units of length if $\kappa =\pm 1$ (then $a(\Delta t)$ is the radius of curvature of the space), otherwise $a(\Delta t)$ may be chosen to be unitless. Subtracting the Friedmann equation from (\ref{ACCE}) gives\smallskip

\begin{equation}  
   \frac{\ddot{a}}{a}
   -
   \left(
      \frac{\dot{a}}{a}
   \right)^2
   =
   -
   4 \pi G
   \cdot
   \left(
      \rho
      +
      \frac{p}{c^2}
   \right)
   +
   \frac{\kappa c^2}{a^2} 
   \;\;\;\;  .
\end{equation}
\smallskip

\noindent To guarantee that $\dot{P} > 0$, this difference must be nonnegative, which can be only if $\kappa \in \{0,1\}$, i.e., the shape of the observable universe is flat or hyperspherical, and the equality\smallskip

\begin{equation} \label{FLUID} 
   p
   =
   -
   \,
   c^2 \rho
   \;\;\;\;    
\end{equation}
\smallskip

\noindent takes place. In such case, $\dot{P}$ turns into\smallskip

\begin{equation}  
   \dot{P}
   =
   \frac{c^3 \Lambda}
   {2 \pi^{2} \hbar G}
   \cdot
   {L_{{_\mathrm{U}}}^4}_{0}
   \cdot
   \left(
      \frac{\kappa c^2}{a^2} \Delta t
      +
      \frac{\dot{a}}{a}
   \right)
   \;\;\;\;  .
\end{equation}
\smallskip

\noindent From the experimental values for critical density, the universe seems to be flat; so, $\kappa = 0$ and $\dot{P}$ is expected to be\smallskip

\begin{equation} \label{SPEED} 
   \dot{P}
   =
   4
   H_{0}
   P_{0}
   \;\;\;\;  ,
\end{equation}
\smallskip

\noindent where $H_{0}$ denotes the Hubble constant, the present value of the Hubble parameter.\bigskip

\noindent Accordingly, the increasing cardinality of the vacuum $P(\Delta t)$ can be thought of as an ideal fluid with negative pressure which acts to push apart gravitationally unbounded objects in the flat universe.\bigskip

\noindent Since the observable universe is vacuum dominated, $\rho$ can be taken approximately equal to $c^{-2} \rho_{\mathrm{vac}}$. Then, substituting (\ref{FLUID}) in (\ref{ACCE}) and recalling that $\rho_{\mathrm{vac}} = -p_{\mathrm{vac}} = \text{const}>0$, one gets the equation\smallskip

\begin{equation}  
   \frac{\ddot{a}}{a}
   =
   -
   \frac{8 \pi G}{3 c^2}
   \cdot
   p_{\mathrm{vac}}
   =
   \text{const}
   >
   0
   \;\;\;\;  .
\end{equation}
\smallskip

\noindent The above means that a constant negative pressure $p_{\mathrm{vac}}$ caused by $\dot{P} > 0$, i.e., the continuously increasing cardinality of the vacuum $P(\Delta t)$, brings about an acceleration in the expansion $a(\Delta t)$.\bigskip

\noindent Typically, to account for a speedup of the expansion $a(\Delta t)$, the idea that space contains \emph{dark energy} – i.e., a hypothetical energy whose the most important property is that it is has negative pressure which is distributed more or less homogeneously – is invoked \cite{Liddle, Peebles, Frieman}. But as it may be inferred from the above, such a speedup may be considered to be a manifestation of continuing growth in the number of points comprising empty space. This growth leads to the runaway expansion of the observable universe where empty space expands, but the amount of matter does not. In consequence, a time may come when the vacuum outweighs everything.\bigskip

\noindent Note that such an outcome differs from \emph{the Big Rip}, a hypothetical cosmological model concerning the fate of the endlessly expanding universe \cite{Caldwell}. Enough just to say that in the Big Rip scenario, all distances in the universe diverge to infinite values, which is impossible if the universe's geometry is finite.\bigskip

\noindent Getting back to the expression for cosmological constant (\ref{LAMB}), it is seen that $\Lambda$ is the ratio between two directly proportional quantities: the size of the observable universe and the size of the universe’s pointset. However, only the former can be determined through the equations of motion (particularly, Einstein’s filed equations). Therefore, nothing but a formalism capable of accounting for both the metric and the pointset of the universe may resolve the “cosmological constant problem”, which is to explain why $\Lambda$ has the specific nonzero value it does.\bigskip

\subsection{A finite geometry of the early universe} 

\noindent Forasmuch as the ratio of $P(\Delta t)$ to $L_{_\mathrm{U}}^{4}(\Delta t)$ remains constant and the expansion of the universe $a(\Delta t)$ is exponential meaning that\smallskip

\begin{equation}  
   L_{_\mathrm{U}}(\Delta t)
   \propto
   \exp{\!\left( H_{0} \Delta t \right)}
   \;\;\;\;  ,
\end{equation}
\smallskip

\noindent one finds\smallskip

\begin{equation} \label{HUBB} 
   P\!\left( \Delta t \right)
   \propto
   \exp{\!\left( 4H_{0} \Delta t \right)}
   \;\;\;\;  .
\end{equation}
\smallskip

\noindent Employing the value $2.19 \times 10^{-18} \,\mathrm{s}^{-1}$ of the Hubble parameter known in 2015 \cite{Planck}, the above can be rewritten as\smallskip

\begin{equation}  
   P\!\left( \Delta t \right)
   \propto
   \exp{\!\left( 0.28 \cdot \Delta T \right)}
   \;\;\;\;  ,
\end{equation}
\smallskip

\noindent where $\Delta T$ is the amount of time whose base unit is billion years. For example, it is predicted that in about 6 billion years the Andromeda - Milky Way collision would occur; so, one can say that by then the cardinality of the vacuum would grow by a factor 5.25.\bigskip

\noindent Changing the sign of $\Delta t$ in the relation (\ref{HUBB}), i.e., extrapolating this relation backwards in time, one can suggest that at the beginning of the observable universe, the size of the pointset of empty space was small, the corollary being that the observable universe had not many points at that moment in time.\bigskip

\noindent This implies that right at the start, a geometry of the observable universe $\mathscr{A}$ might be characterized by such prominent finiteness that even with limited computational resources the cardinalities of every pointset and its proper subset in $\mathscr{A}$ could be distinguished from each other. Correspondingly, the distance function $d$ on $\mathscr{A}$ would map $\mathcal{P}_{\!\mathscr{A}} \times \mathcal{P}_{\!\mathscr{A}}$ to a field that contains a conspicuously finite number of elements, namely, $d\!: \mathcal{P}_{\!\mathscr{A}} \times \mathcal{P}_{\!\mathscr{A}} \to \mathbb{F}$. However, the problem is that the said function violates the metric axiom $\mathbf{M3}$ which asserts that the distance between distinct points must be non-zero.\bigskip

\noindent To show that, assume that the field $\mathbb{F}$ has just four elements, specifically, $\mathbb{F}_{4}=\{0,1,\alpha,1+\alpha\}$, such that $\alpha^2=1+\alpha$, $1 \times \alpha=\alpha\times 1=\alpha$, and $x+x=x-x=0$ (i.e., subtraction is identical to addition). Consider a plane over $\mathbb{F}_{4}$ and two points on this plane, namely, $A=(1,1)$ and $B=(\alpha,\alpha)$. Then,\smallskip

\begin{equation}  
   d^{\,2}(A,B)
   =
   (1-\alpha)^2
   +
   (1-\alpha)^2
   =
   \alpha
   +
   \alpha
   =
   0
   \;\;\;\;  .
\end{equation}
\smallskip

\noindent According to the table of multiplication in $\mathbb{F}_{4}$ (see for example \cite{Mullen}), $x^2=0$ only if $x=0$; hence, $d(A,B)=0$ even though $A \neq B$. Subsequently, one may conclude that in a conspicuously finite geometry $\mathscr{A}$ (i.e., a geometry whose pointset is small), there is at least one pair of distinct points which are not space separated from each other. In symbols,\smallskip

\begin{equation}  
   \bigg(
   \mathrm{card}\left( \mathcal{P}_{\mathscr{A}}\right)
   =
   q^{\mathrm{dim}(\mathscr{A})}
   \text{ is small}
   \bigg)
   \to
   \bigg(
   \exists x,y \in \mathcal{P}_{\!\mathscr{A}}
      \,\Big( x \neq y \,\sqcap\, d(x,y) = 0 \Big)
   \bigg)
    \;\;\;\;  .
\end{equation}
\smallskip

\noindent Consider the slope of a function $f(x)$ at the point $A$ in $\mathscr{A}$. Due to the breakdown of the metric axiom $\mathbf{M3}$, one can no longer pick the second point $B$ in $\mathscr{A}$ and compute the slope of the secant line passing through these two points such that\smallskip

\begin{equation}  
   \text{slope}
   =
   \frac{f(A) - f(B)}{d(A,B)}
    \;\;\;\;  .
\end{equation}
\smallskip

\noindent Clearly, if $d(A,B)=0$ and $A \neq B$, the secant line will not exist. Since the concept of the slope of a function at a particular point is essential to differential calculus, one can state that the differential equations of motion including the Einstein field equations for gravitation (which provide the expression for the first derivative of the cardinality $P(t)$ with respect to time $t$) do not hold true around the beginning of the observable universe.\bigskip

\noindent Evidently, a conspicuously finite geometry $\mathscr{A}$, which characterizes the observable universe at an early stage, does not have to be metric (i.e., one that is based on \emph{all} the metric axioms and thence determines the size of geometrical magnitudes such as lengths, areas, and volumes). Even more interesting, $\mathscr{A}$ may have no metric space realization at all. This means that $\mathscr{A}$ could not  be realizable by points and lines (or more generally, geodesics) of any metric space. To elucidate this regard, let us introduce $\mathcal{L}_{\mathscr{A}}$, the set of lines that consists of subsets of $\mathcal{P}_{\!\mathscr{A}}$, and $I_{\mathscr{A}} \subseteq \mathcal{P}_{\!\mathscr{A}} \times \mathcal{L}_{\mathscr{A}}$, the incidence relation between $\mathcal{P}_{\!\mathscr{A}}$ and $\mathcal{L}_{\mathscr{A}}$ (that concerns which points lie on which lines). Consider the geometry $\mathscr{A}$ whose \emph{incidence structure}, that is, the triple $(\mathcal{P}_{\!\mathscr{A}}, \mathcal{L}_{\mathscr{A}}, I_{\mathscr{A}})$, realizes the Hesse configuration \cite{Gropp} in which every line $l \in \mathcal{L}_{\mathscr{A}}$ through two points $x,y \in \mathcal{P}_{\!\mathscr{A}}$ contains a third point $z \in \mathcal{P}_{\!\mathscr{A}}$. By contrast, according to Sylvester–Gallai theorem \cite{Elkies}, in each metric space, all finite incidence structures $(\mathcal{P}, \mathcal{L}, I)$ contain pointsets $\mathcal{P}$ that are either colinear or include a pair of points whose geodesic cannot have one more point of $\mathcal{P}$.\bigskip

\noindent The preceding paragraph may be understood to imply that the first stages of the observable universe can in no way be visualized by means of metric geometries. To give an example, it is customary to picture the observable universe’s initial state as one where any two things in the present day universe would be located arbitrarily close to each other (which may mean that at the beginning, the observable universe was infinitesimally small in size). But, as the last section demonstrates, that picture might be wrong since the conception of distance may turn out to be senseless during the earliest moments of cosmic time.\bigskip

\noindent One more example, the classical version of the Big Bang cosmological model of the universe contains a singularity at $t=0$, where all timelike geodesics have no extensions into the past. A key tool used for the visualization of this model is a timelike geodesic between points $A$ and $B$ which represent an arbitrary event in the early phase of the Big Bang and the initial gravitational singularity, in the order named. Most importantly, the geodesic cannot be extended in a smooth manner past the point $B$ to some point $C$ (depicting what might come “out” of the initial singularity in the past) because, e.g., spacetime is punctured at the point $B$ or looks like a cone around this point. Then again, if the incidence structure of the observable universe during its earliest moments was Hesse-like, then such a visualization would be wrong. In that case, there would be at least 3 lines through each point, therefore, the line passing through the point $A$ would unabruptly (i.e., without a jump) reach the point $C$, even if the point $B$ (and its incident lines) had been removed from the Hesse configuration. That is, the notion of geodesic incompleteness – and thus the notion of singularity – cannot take place in a Hesse-like geometry.\bigskip

\noindent For all that, in this geometry, it is not the case that a certain point contained in each line $l \in \mathcal{L}_{\mathscr{A}}$ is between two other points lying on the same line. In other words, because the ordering of the points lying on each line is irrelevant in $\mathscr{A}$, one cannot give meaning to the concept of “betweenness” therein. This implies that given any three discrete events connected by a line $l \in \mathcal{L}_{\mathscr{A}}$, one would be unable to say which event happened between which events. Thereupon it may be argued that time as a collection of events wherein order matters is an emergent phenomenon: Time did not exist until the cardinality of the observable universe became vast.\bigskip

\subsection{A finite geometry of spacetime on very small scales} 

\noindent Consider a proper subset $\mathcal{P}_{\!\mathscr{U}}$ of a vast pointset $\mathcal{P}_{\!\mathscr{A}}$. Let $\mathcal{P}_{\!\mathscr{U}}$ be the pointset of some subspace $V_{\mathscr{U}}$ in $\mathscr{A}$ whose cardinality is $\mathrm{card}\left( \mathcal{P}_{\mathscr{U}}\right) = \mathrm{card}\left( \mathbb{F}_{\mathscr{U}} \right)^{\mathrm{dim}(\mathscr{A})}$. The size of $V_{\mathscr{U}}$ can be estimated as the metric diameter of $V_{\mathscr{U}}$, i.e., the least upper bound of the set of all distances between pairs of points in $\mathcal{P}_{\!\mathscr{U}}$:\smallskip

\begin{equation} \label{DIAM} 
   \mathrm{diam}\!\left( V_{\mathscr{U}} \!\right)
   =
   \mathrm{sup}_{\,x,y \,\in\, \mathcal{P}_{\!\mathscr{U}} \,\sqcap\, x \neq y}
   \!\left\{
      d(x,y)
   \right\}
   \;\;\;\;  .
\end{equation}
\smallskip

\noindent At the same time, $\mathrm{diam}\!\left( V_{\mathscr{U}} \!\right)$ is a line in $\mathscr{A}$ that contains $\mathrm{card}\left( \mathbb{F}_{\mathscr{U}} \right)$ points of $\mathcal{P}_{\!\mathscr{U}}$. Therefore, $\mathrm{diam}\!\left( V_{\mathscr{U}} \!\right)$ can be presented as\smallskip

\begin{equation} \label{DDIST} 
   \mathrm{diam}\!\left( V_{\mathscr{U}} \!\right)
   =
   d \left( X_{1}, X_{\mathrm{card}\left( \mathbb{F}_{\mathscr{U}}\right)} \right)
   =
   \sum_{n=1}^{\mathrm{card}\left( \mathbb{F}_{\mathscr{U}} \right) - 1}
      d \left( X_{n}, X_{n+1} \right)
   \;\;\;\;  ,
\end{equation}
\smallskip

\noindent where all $X_{n}$ are in $\mathcal{P}_{\!\mathscr{U}}$.\bigskip

\noindent The smaller is the field $\mathbb{F}_{\mathscr{U}}$, the smaller is the metric diameter of the subspace $V_{\mathscr{U}}$. To make the relation between $\mathrm{card}\left( \mathbb{F}_{\mathscr{U}} \right)$ and $\mathrm{diam}\!\left( V_{\mathscr{U}} \!\right)$ explicit, assume that the points on the line $\mathrm{diam}\!\left( V_{\mathscr{U}} \!\right)$ are at the same distance from each other. Then, the expression (\ref{DDIST}) can be rewritten as\smallskip

\begin{equation} \label{DREL} 
   \mathrm{diam}\!\left( V_{\mathscr{U}} \!\right)
   =
   d_0  
   \left(
   \mathrm{card}\!\left( \mathbb{F}_{\mathscr{U}} \!\right) - 1
   \right)
   \;\;\;\;  ,
\end{equation}
\smallskip

\noindent where $d_0$ is the constant of proportionality that can be chosen arbitrarily small.\bigskip

\noindent It may be tempting to think that the above relation equally applies for any field $\mathbb{F}_{\mathscr{U}}$, down to the situation where $\mathbb{F}_{\mathscr{U}}$ has just a single point. In that case, a macroscopically sized subspace $V_{\mathscr{U}}$ would correspond to a vast field $\mathbb{F}_{\mathscr{U}}$, whereas a very small metric diameter of $V_{\mathscr{U}}$ would match up with a conspicuously finite $\mathbb{F}_{\mathscr{U}}$.\bigskip

\noindent However, it was not. As it has been demonstrated on the case of the early universe, when $\mathbb{F}_{\mathscr{U}}$ becomes conspicuously finite, say, $\mathrm{card}\left( \mathbb{F}_{\mathscr{U}} \right) = 4$, the metric axiom $\mathbf{M3}$ breaks down meaning that the Eq.(\ref{DIAM}) and the Eq.(\ref{DDIST}) get nonsensical, not to mention the relation (\ref{DREL}).\bigskip

\noindent Let us estimate $\mathrm{diam}\!\left( V_{\mathscr{U}} \!\right)_{\mathrm{min}}$, the lowest possible value of a metric diameter, below which the formula (\ref{DDIST}) turns into nonsensical. Towards this end, let us bring up the pointset density in the observable universe denoted by $\rho_{\!{_P}}\!(t)$ and given by\smallskip

\begin{equation}  
   \rho_{\!{_P}}\!(t)
   =
   \frac{P\!\left( t \right)}{{L_{_\mathrm{U}}^3}\!\left( t \right)}
   \;\;\;\;  .
\end{equation}
\smallskip

\noindent Subsequently, a box of the vacuum would contain only a few points if its volume were about $\rho_{\!{_P}}^{-1}(t)$. Providing  $\mathrm{diam}^3\left( V_{\mathscr{U}} \!\right)_{\mathrm{min}}$ coincides with that volume, one finds\smallskip

\begin{equation}  
   \mathrm{diam}\!\left( V_{\mathscr{U}} \!\right)_{\mathrm{min}}
   \sim
   \sqrt[3]
   {
      \frac{\pi \hbar c}{ \rho_{\mathrm{vac}} \cdot L_{_\mathrm{U}}\!\left( t \right) }
   }
   \;\;\;\;  .
\end{equation}
\smallskip

\noindent Substituting $\rho_{\mathrm{vac}}$ and $L_{_\mathrm{U}}\!\left( t \right)$ obtained from the recent cosmological observations defines $\mathrm{diam}\!\left( V_{\mathscr{U}} \!\right)_{\mathrm{min}}$ at the present time as $\sim 5.9 \times 10^{-15} \,\mathrm{m} = 5.9 \,\mathrm{fm} $.\bigskip

\noindent Let us note in passing, that if the observable universe were to be describable within the limits of Zermelo-Fraenkel set theory, then the value of the vacuum energy density (guessed by dimensional analysis as $\rho_{\mathrm{vac}} = c^4 (8 \pi G)^{-1} \ell_{P}^{-2}$ from the premise that an effective local quantum field theory cannot hold on lengths $\ell$ shorter than the Planck scale $\ell_{P} \sim 10^{-35} \,\mathrm{m}$  \cite{Carroll}) would be $\sim 1.85 \times 10^{112} \,\mathrm{J}\cdot \,\mathrm{m}^{-3}$. That would settle $\mathrm{diam}\!\left( V_{\mathscr{U}} \!\right)_{\mathrm{min}}$ on the value $\sim 1.83 \times 10^{-55}  \,\mathrm{m}$. However, not only does this value seem unreasonable, but – what is more important – it contradicts the premise $\ell \notle \ell_{P}$.\bigskip

\noindent This implies that a metric space represented by a vast pointset $\mathcal{P}_{\!\mathscr{A}}$ cannot have an arbitrarily small subspace, specifically, one whose diameter is on par with (or smaller than) the length scale for the strong interactions $\ell_{s} \sim 10^{-15} \,\mathrm{m}$. As a result, differential equations of motion including equations for gravitation that contain derivatives cannot be applicable at the length scale which is close to or shorter than $\ell_{s}$. That is to say, the geometry of spacetime seen as metric at macroscopic scales ceases to be such on scales $\apprle \ell_{s}$ whereon both the notion of distance and the notion of time turn into unsuitable.\bigskip

\noindent A quick explanation of the unexpectedly high cut between metric and non-metric geometries (around the length of a few femtometers as opposed to being, say, a few $\ell_{P}$) is in order.\bigskip

\noindent Recall that a path taken by a freely moving (e.g., under the influence of gravity alone) point-like particle is called a geodesic trajectory. Given the field $\mathbb{F}_{\mathscr{U}}$, a trajectory within the subspace $V_{\mathscr{U}}$ can be formulated as a set $T$ of points in $\mathcal{P}_{\mathscr{U}}$, in particular:\smallskip

\begin{equation}  
   T
   =
   \bigg\{
      \Big(
         x(t),y(t),z(t)
      \Big)
      \!\in \mathbb{F}_{\mathscr{U}}^3
      \,\bigg|\,
      t \in \left[ t_i, t_f\right] \subset \mathbb{F}_{\mathscr{U}}
      \sqcap
      t_i < t_f
      \bigg.
   \bigg\}
   \;\;\;\;  ,
\end{equation}
\smallskip

\noindent where $x(t)$, $y(t)$, and $z(t)$ are smooth functions, i.e., functions that have the slope at each point in $\mathcal{P}_{\mathscr{U}}$. However, as it follows from what has been said before, that may not be the case at the length scale $\apprle \ell_{s}$ meaning that the set $T$ may not exist on lengths $\apprle \ell_{s}$. This entails that the classical notion of trajectory breaks off at the length scale $\apprle \ell_{s}$.\bigskip

\noindent On the other hand, within a Euclidean or traditional non-Euclidean geometry (a metric geometry with an infinite pointset), Heisenberg's uncertainty principle implies a significant departure from the classical notion of trajectory for a point-like particle (such as an electron, quark, or photon) moving freely over distances compatible with the length scale $\ell_{s}$. Hence, a conspicuously finite geometry whose incidence structure has no metric realization can be comprehended as a geometric account of quantum behavior, that is, the behavior of nature at the length scale $\ell_{s}$ whereupon quantum fluctuations within the vacuum of space start to get measurable consequences.\bigskip\\

\noindent It appears as if the core idea of general relativity that the metric is determined by the matter and energy content of spacetime is not enough to capture the geometry of the observable universe: A true TOE must also give reasons for the universe’s pointset as well as for the universe’s incidence structure.\bigskip

\bibliographystyle{References}
\bibliography{Finite_physics_Ref}

\end{document}